\documentclass[prd,aps,floatfix,preprint,nofootinbib]{revtex4}
\usepackage{float} 
\usepackage{graphicx,color}
\usepackage{epstopdf}
\usepackage{amssymb}
\usepackage{amsmath}
\usepackage{subfigure}
\usepackage{epsfig}
\usepackage{array}
\usepackage{hyperref}

\def\simge{\mathrel{%
       \rlap{\raise 0.511ex \hbox{$>$}}{\lower 0.511ex \hbox{$\sim$}}}}
\def\simle{\mathrel{
       \rlap{\raise 0.511ex \hbox{$<$}}{\lower 0.511ex \hbox{$\sim$}}}}

\newcommand{\figcaption}[1]{\def\@captype{figure}\caption{#1}}
\newcommand{\tblcaption}[1]{\def\@captype{table}\caption{#1}}
\newcommand{\no}{\nonumber}
\newcommand{\cpn}{CP$^{N\!-\!1}$\ }
\newcommand*{\doi}[1]{\href{https://doi.org/#1}{doi:\ #1}}

\newcommand{\SU}{\mathrm{SU}}


\begin{document}

\title{Is $N=2$ Large?}

\author{Ryuichiro Kitano$^{1,2}$}
\email{ryuichiro.kitano@kek.jp}
\author{Norikazu Yamada$^{1,2}$}
\email{norikazu.yamada@kek.jp}
\author{Masahito Yamazaki$^3$}
\email{masahito.yamazaki@ipmu.jp}
\affiliation{
   $^1$ High Energy Accelerator Research Organization (KEK), %
        Tsukuba 305-0801, Japan\\
   $^2$ Graduate University for Advanced Studies (SOKENDAI), %
        Tsukuba 305-0801, Japan\\
   $^3$ Kavli Institute for the Physics and Mathematics of the Universe
	(WPI), University of Tokyo, Kashiwa, Chiba 277-8583, Japan
}
\date{\today}

\begin{abstract}
 We study $\theta$ dependence of the vacuum energy for the 4d $\SU(2)$
 pure Yang-Mills theory by lattice numerical simulations. The response
 of topological excitations to the smearing procedure is investigated in
 detail, in order to extract topological information from smeared gauge
 configurations. We determine the first two coefficients in the $\theta$
 expansion of the vacuum energy, the topological susceptibility $\chi$
 and the first dimensionless coefficient $b_2$, in the continuum limit.
 We find consistency of the $\SU(2)$ results with the large $N$ scaling.
 By analytic continuing the number of colors, $N$, to non-integer values,
 we infer the phase diagram of the vacuum structure of  $\SU(N)$ gauge theory
 as a function of $N$ and $\theta$.
 Based on the numerical results, we provide quantitative evidence that 4d
 $\SU(2)$ Yang-Mills theory at $\theta = \pi$ is gapped with spontaneous
 breaking of the CP symmetry.
 
\end{abstract}
\maketitle
\tableofcontents 

\section{Introduction}\label{sec:intrroduction}

The $\theta$ term of the Yang-Mills theory
determines how to weight different topological sectors in the path integral.
Since the $\theta$ parameter is the coefficient of a total derivative term in the Lagrangian,
the $\theta$-dependences of observables can be explored only through non-perturbative methods.

The special value $\theta=\pi$ has been of particular interest.
In the classic literature~\cite{Witten:1980sp,tHooft:1981bkw,Witten:1998uka}, spontaneous CP
violation of the 4d SU($N$) Yang-Mills theory at $\theta=\pi$ was demonstrated in the large $N$ limit~\cite{tHooft:1973alw}.
More recently, an anomaly matching argument involving generalized global symmetries~\cite{Gaiotto:2014kfa}
showed that the CP symmetry in the confining phase has to be broken even at finite $N$~\cite{Gaiotto:2017yup}.
A similar conclusion was derived by studying restoration of the equivalence
of local observables between SU($N$) and SU($N$)/$Z_N$ gauge theories in the infinite volume limit~\cite{Kitano:2017jng}.
See, for example,
Refs.~\cite{Azcoiti:2003ai,Yamazaki:2017dra,Wan:2018zql} for other approaches.

While lattice numerical simulations are ideal tools to
explore non-perturbative dynamics of gauge theories, direct
simulations at $\theta=\pi$ has been challenging due to the notorious sign
problem.\footnote{Recent and related developments towards direct simulations are found,
for example, in Refs.~\cite{Hirasawa:2020bnl,Gattringer:2020mbf,Sulejmanpasic:2020lyq}.}
Nevertheless, lattice simulations have been successfully used to
determine the first few coefficients in the $\theta$ expansion of the
vacuum energy for finite $N$. 
On the one hand, below the critical temperature
$T_c$ these coefficients turn out to be consistent with the large $N$ scaling down to $N=3$~\cite{Lucini:2001ej,DelDebbio:2002xa,Bonati:2016tvi},
which indicates spontaneous CP violation and the discontinuity of the vacuum energy across $\theta=\pi$.
On the other hand, above $T_c$ the coefficients determined at $N=3$ and $6$ are found
to be consistent with the dilute instanton gas approximation (DIGA)
~\cite{Bonati:2013tt}, which predicts continuous behavior for the
vacuum energy across $\theta=\pi$.

The \cpn model in two dimensions shares many non-perturbative properties
with the four-dimensional SU($N$) Yang-Mills
theory~\cite{Eichenherr:1978qa,DAdda:1978vbw}, and hence
provides useful insights into the latter.
For $N\ge 3$, the model is believed to show spontaneous CP violation at
$\theta=\pi$~\cite{Beard:2004jr}.
By contrast the case with $N=2$ is believed to be special and argued to
become gapless at $\theta=\pi$ with unbroken CP
symmetry~\cite{Haldane:1982rj,Haldane:1983ru,Affleck:1987ch,Shankar:1989ee,Affleck:1991tj,Zamolodchikov:1992zr,Bietenholz:1995zk,Alles:2007br,Alles:2014tta}.
Motivated by similarities between the 4d Yang-Mills theory and the 2d \cpn model,
it is natural to ask if the 4d SU($N$) Yang-Mills theory at $\theta=\pi$ shows distinctive 
behavior for small values of $N$, such as $N=2$.\footnote{The $\mathbb{Z}_N$ subgroup of the flavor symmetry of the 2d \cpn model can be regarded as a counterpart of the 
1-form $\mathbb{Z}_N$ center symmetry of the 4d $\SU(N)$ pure Yang-Mills theory.}

In this work we explore the $\theta$ dependence of the vacuum
energy of the 4d SU(2) pure Yang-Mills gauge theory.
In sec.~\ref{sec:lattce-simulations}, we perform lattice numerical
calculations to determine the first two coefficients in the $\theta$
expansion of the vacuum energy. 
The response of topological excitations to the smearing procedure is investigated in detail,
in order to efficiently extract physical information from lattice
configurations.
The coefficients determined for $N=2$ are compared to those previously
obtained for $N\ge 3$, to see whether the result at $N=2$ can be seen as
a natural extrapolation of those for $N\ge 3$.
In sec.~\ref{sec:discussion}, we begin with theoretical arguments for 
different behaviors of 4d $\SU(N)$ theory, for large $N$  and for small $N$
as we analytically continue the values of $N$.
We then interpret the numerical results of sec.~\ref{sec:lattce-simulations}
and provide quantitative evidence that the 4d $\SU(2)$ theory belongs to the 
``large $N$'' class, and is gapped and has spontaneous breaking of CP
symmetry at $\theta=\pi$.

\section{Lattice Simulations}\label{sec:lattce-simulations} 

The vacuum energy can be expanded around $\theta=0$ as
\begin{align}\label{E_expand}
   E(\theta)-E(0)
&= \frac{\chi}{2}\theta^2
   \left(1 + b_2\,\theta^2 + b_4\,\theta^4 + \cdots \right)\ ,
\end{align}
where $\chi$ is the topological susceptibility, and $b_{2i}$
($i=1, 2, 3, \cdots$) are dimensionless coefficients describing the
deviation of the topological charge distribution from the Gaussian.
These quantities can be determined from the lattice configurations generated at $\theta=0$ as
\begin{align}
& \chi = \frac{\langle Q^2 \rangle_{\theta=0}}{V}\ ,\\
& b_2=-\frac{      \langle Q^4 \rangle_{\theta=0}
              - 3\,\langle Q^2 \rangle_{\theta=0}^2}
            {12\,\langle Q^2 \rangle_{\theta=0}}
\ ,\\
& b_4= \frac{       \langle Q^6 \rangle_{\theta=0}
              - 15\,\langle Q^2 \rangle_{\theta=0}\,
                  \langle Q^4 \rangle_{\theta=0}
              + 30\,\langle Q^2 \rangle_{\theta=0}^3}
            {360\,\langle Q^2 \rangle_{\theta=0}}
\ ,
\end{align}
where $Q$ is the topological charge, whose precise definition is given
in eqs.~(\ref{eq:q})-(\ref{eq:q-coefs}), and
$\langle\cdots\rangle_{\theta=0}$ denotes an ensemble average over
configurations generated at $\theta=0$.
According to the large $N$ analysis~\cite{Witten:1980sp,Witten:1998uka},
these quantities can be expressed, as a function of $N$, as
\begin{align}
 \chi(N) &= \chi(\infty) + O(N^{-2})\ ,\\
b_{2i}(N)&= \frac{b_{2i}^{(1)}}{N^{2i}}+O\left(\frac{1}{N^{2i+2}}\right)\ . \label{b_N}
\end{align}
By contrast the dilute instanton gas approximation leads to
$E(\theta)-E(0)=\chi(1-\cos\theta)$,
and hence the coefficients, $b_2^{\rm DIGA}=-1/12$,
$b_4^{\rm DIGA}=1/360$, $\cdots$, are completely determined.
We attempted calculating $b_4$ as well as $\chi$ and $b_2$.
We could obtain only a loose bound $-0.1<b_4<0.1$ due to a large
statistical uncertainty. In the following, we focus on the determinations of $\chi$ and $b_2$.

\subsection{Lattice Setup}
\label{subsec:setup}

The SU(2) gauge action on the lattice is described as
\begin{align}
 S_g = 6\,\beta\,N_{\rm site}\,\left\{c_0(1-W_P)+2\,c_1(1-W_R)\right\}\ ,
\end{align}
where $\beta=4/g^2$ is the lattice gauge coupling, $W_P$ and $W_R$
are the $1\times 1$ plaquette and the $1\times 2$ rectangle averaged over
four dimensional lattice sites, respectively, and $c_0$ and $c_1$ satisfying $c_0=1-8c_1$ are the improvement
coefficients.
We take the tree-level Symanzik improved action~\cite{Weisz:1982zw},
which is realized by $c_1=-1/12$.
To investigate the continuum limit, three values of the lattice
coupling ($\beta$=1.750, 1.850 and 1.975) are taken.
The lattice size is $N_{\rm site}=N_S^3\times N_T$ with $N_S=16$ and
$N_T=2\times N_S$.
We also perform simulations with $N_S=24$ on our finest lattice to
check finite volume effects.
The lattice spacing at each $\beta$ is taken from
$N_{T_c}=1/(a(\beta)T_c)$ obtained in Ref.~\cite{Giudice:2017dor}, where
$T_c$ is the critical temperature for the $\SU(2)$ pure Yang-Mills
theory.
The value $N_{T_c}$ is then transformed to $(aT_c)^2$ for later use.
To have an intuition about how large our lattice is, we estimated
$L\,\sigma_{\rm str}^{1/2}$ at each lattice, using
$T_c/\sqrt{\sigma_{\rm str}}=0.7091(36)$~\cite{Lucini:2003zr}, where $L$
denotes the physical length of the spatial direction, {\it i.e.}
$L=a\,N_S$, and $\sigma_{\rm str}$ is the representative dynamical scale
(the sting tension).
Gauge configurations are generated by hybrid Monte Carlo method and
are stored every 10 trajectories.
Simulation parameters including the lattice spacings, the lattice size
and the number of configurations (denoted as statistics in the table)
are summarized in Tab.~\ref{tab:parameters}.
\begin{table}[tbp]
 \begin{center}
 \begin{tabular}{|p{6ex} p{6ex}|p{6ex}p{10ex}p{8ex}r|}
 \hline
$\beta$ & $N_S$ & $N_{T_c}$ & $(a\,T_c)^2$ &
  $L\,\sigma_{\rm str}^{1/2}$ & statistics 
\\
 \hline
  1.750 & 16 & 4.65 & 0.0462 & 4.9 &  80,100 \\
  1.850 & 16 & 6.50 & 0.0237 & 3.5 &  71,040 \\
  1.975 & 16 & 9.50 & 0.0111 & 2.4 &  30,490 \\
  1.975 & 24 & 9.50 & 0.0111 & 3.6 & 131,830 \\
 \hline
 \end{tabular}
 \caption{
  The simulation parameters.
  $T_c$ denotes the critical temperature.
  $N_{T_c}$ is determined from Ref.~\cite{Giudice:2017dor}.
  The uncertainties of $N_{T_c}$ are below 1\% and hence neglected in
  the following.}
 \label{tab:parameters}
 \end{center}
\end{table}

\subsection{Smearing and Definition of Topological Charges on the Lattice}
\label{subsec:smearing-topological-charge}

Among several equivalent methods often used in the
literature~\cite{Bonati:2014tqa,Alexandrou:2015yba,Alexandrou:2017hqw},
we choose the combination of the APE smearing~\cite{Albanese:1987ds} and
the 5-loop improved operator~\cite{deForcrand:1997esx} to calculate
topological charge on each configuration.
Topological charges on the lattice are obscured by short distance
fluctuations, which we remove by introducing a smoothing technique.
In the APE smearing, new link variables $U_\mu^{\rm (new)}$ are
constructed from old ones $U_\mu^{\rm (old)}$ as
\begin{align}
  U_\mu^{\rm (new)}
&=  \mathrm{Proj}
  \left[
  (1-\rho) U_\mu^{\rm (old)}(x) + \rho\,X_\mu(x)
  \right]\ ,
\\
  X_\mu(x)
&= \sum_{\nu\ne\mu}
  \left[
     U_\nu^{\rm (old)}       (x        )
     U_\mu^{\rm (old)}       (x+\hat\nu)
     {U_\nu^{\rm (old)}}^\dag(x+\hat\mu) \right.
\no\\&\qquad\left.
   + {U_\nu^{\rm (old)}}^\dag(x-\hat\nu)
     U_\mu^{\rm (old)}       (x-\hat\nu)
     U_\nu^{\rm (old)}       (x-\hat\nu+\hat\mu)
  \right]\ ,
\end{align}
where $\rm{Proj}$ acts as the projection back to an $\SU(2)$ element.
This procedure minimizes the action density.
The parameter $\rho$ is taken to be 0.2, which corresponds to
$\alpha_{\rm APE}=6 \rho/(1+5\,\rho)=0.6$ in
Ref.~\cite{Alexandrou:2017hqw}.

The 5-loop improved topological charge operator is given by
\begin{align}
&Q=\sum_x q(x)\ ,
\label{eq:q}\\
&q(x)=\sum_i c_i\, q_{m_i,n_i}(x)\ ,
\label{eq:qdensity}\\
&q_{m,n}(x)=\frac{1}{32\pi^2}\frac{1}{m^2n^2}
\sum_{\mu,\nu,\rho,\sigma}\epsilon_{\mu,\nu,\rho,\sigma}
{\rm Tr}\left[
 \hat{F}_{\mu,\nu}(x;m,n)\hat{F}_{\rho,\sigma}(x;m,n)
 \right]\ ,\\
&\hat{F}_{\mu,\nu}(x;m,n)=
\frac{1}{8}\,{\rm Im}\left[
 \big\{\mbox{oriented clover average of ($m\times n$) plaquette}\big\}
+ \big\{\mbox{($m \leftrightarrow n$)}\big\}
 \right]\ ,
\end{align}
where $(m_i,n_i) = (1,1), (2,2), (1,2), (1,3), (3,3)$ for $i = 1,\dots ,5$ and the coefficients are given by
\begin{align}
\begin{split}
&c_1=(19-55 c_5)/9 ,\quad
 c_2=(1-64 c_5)/9,\quad
 c_3=(-64+640 c_5)/45 ,\\
&c_4=1/5-2 c_5 ,\quad
 c_5=1/20 \ .
 \end{split}
\label{eq:q-coefs}
\end{align}
This operator is free of $O(a^2)$ and $O(a^4)$ terms.
The replacement above is done once for all
link variables, for each step of the smearing.
The smearing is carried out every 10 trajectories, and the topological
charge is measured after every smearing step.

\subsection{Response to Smearing}
\label{subsec:response-to-smearing}

As mentioned above, the smearing is introduced to remove short distance
fluctuations, which distort physical topological excitations through
local lumps with the size of the lattice spacing.
The measurement of the topological charge is therefore reliable only after a
suitable number of smearing steps.
However, the smearing may also affect physical topological excitations.
The previous dedicated studies revealed that the smearing induces
pair-annihilation, ``melting away'' or ``falling through the
lattice''~\cite{Laursen:1990ec,deForcrand:1997esx,BilsonThompson:2004ez}.
In Ref.~\cite{BilsonThompson:2004ez}, it was found that topological
objects go through several characteristic phases during the cooling
procedure.
In the first phase, the size of topological objects grows with the
cooling, and some of them eventually melt away and some
pair-annihilate.
Then, the second phase comes where only relatively slow shrinkage of the
objects takes place and eventually they disappear after long enough
cooling.
Assuming that the similar phases show up in the procedure of APE
smearing, we will in the following determine the boundary between the two phases.

\begin{figure}[tbp]
 \begin{center}
  \begin{tabular}{cc}
  \includegraphics[width=0.51 \textwidth]
  {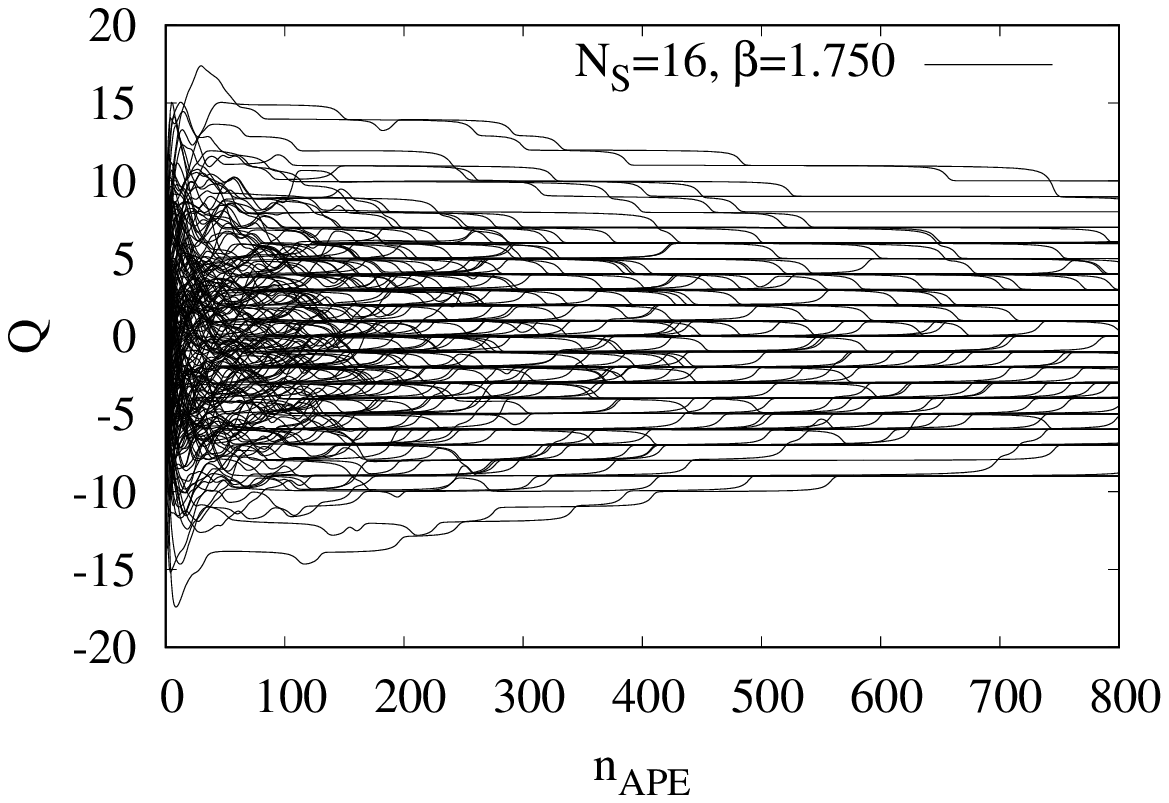} &
  \includegraphics[width=0.51 \textwidth]
  {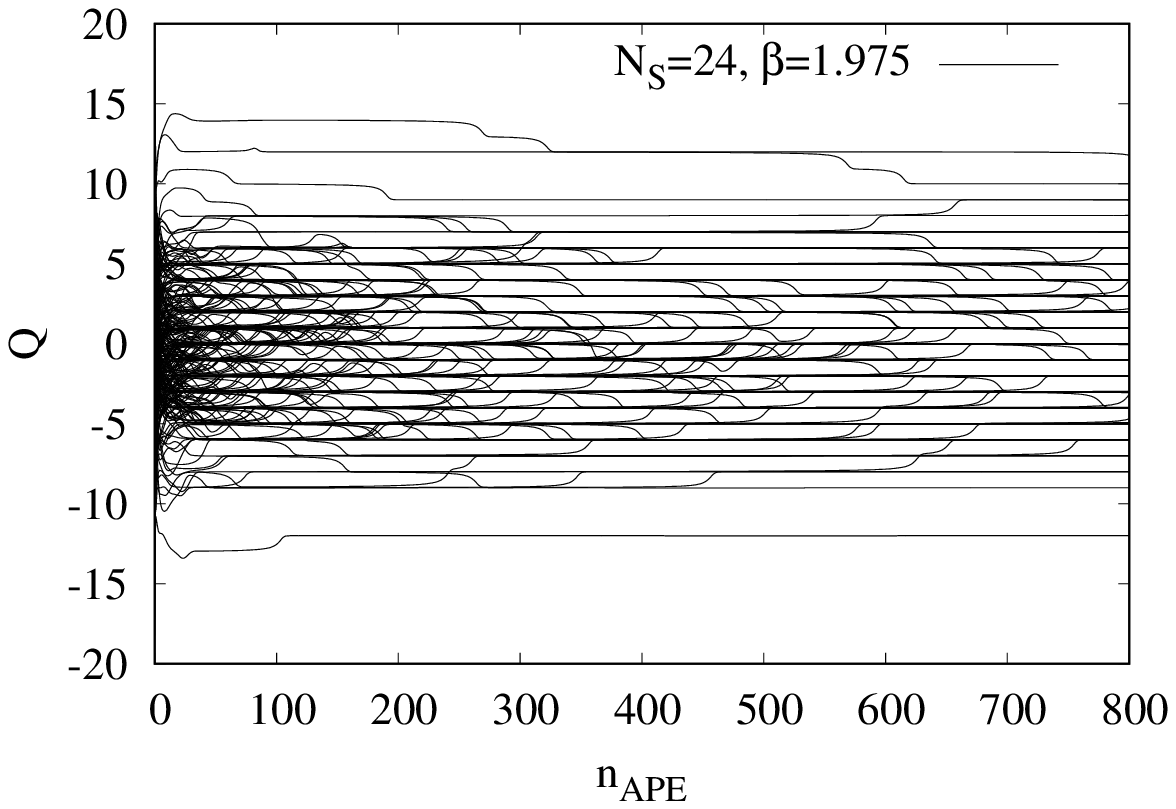}
  \end{tabular}
 \end{center}
 \caption{
 Smearing history of topological charge $Q$ as a function of the
 number of smearing steps $n_{\rm APE}$ for randomly chosen 200
 configurations at $\beta=1.750$ and 1.975.}
 \label{fig:nape_vs_q}
\end{figure}

In order to explore how the smearing changes topological properties, we
first look at the smearing history of the topological charge $Q$ as a
function of the smearing steps $n_{\rm APE}$. 
Fig.~\ref{fig:nape_vs_q} shows the history obtained at $\beta=1.750$
and 1.975.
At relatively small $n_{\rm APE}$, $Q$ changes frequently, and most of
the changes here are expected to be associated with the removal of short
distance fluctuations.
We deduce that this range of $n_{\rm APE}$ corresponds to the first
phase.
The frequency of change in $Q$ is somewhat reduced as $n_{\rm APE}$ and
$\beta$ increase, but the change steadily continues.
In this region, both the increase and the decrease of $Q$ happen mostly
by one unit, and a change takes $O(10)$ steps to be completed.
This range of $n_{\rm APE}$ is identified with the second phase.
In the following, we discuss quantitative differences between the
two phases.

\begin{figure}[phtb]
 \begin{center}
 \vspace*{-3ex}
  \begin{tabular}{c}
  \includegraphics[width=0.8 \textwidth]
  {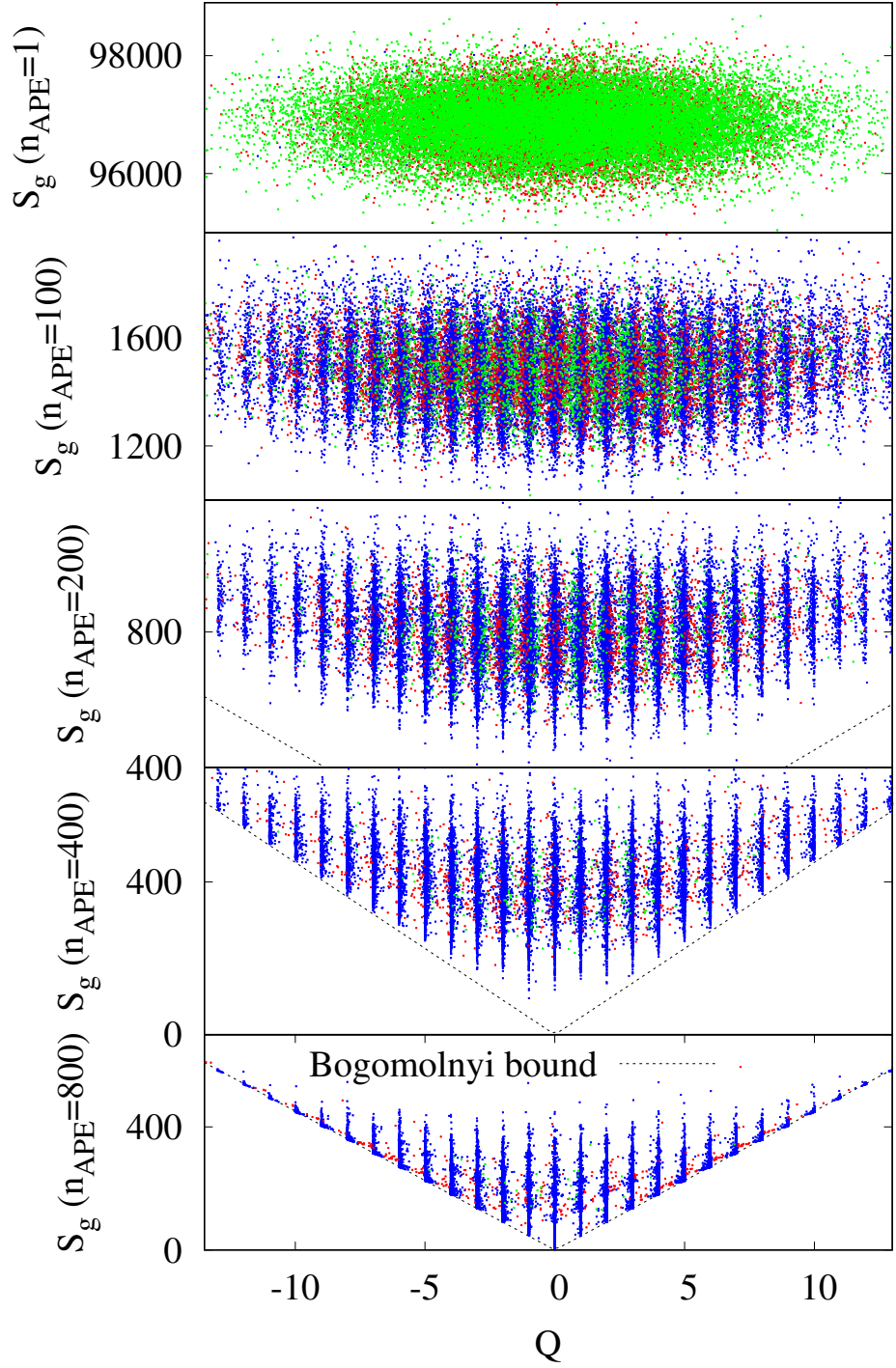}
  \end{tabular}
 \vspace{5ex}
 \end{center}
 \caption{Correlation between the topological charge $Q$ and the action
 $S_g$ at smearing steps $n_{\rm APE}=1,100,200,400,800$ (from top to
 bottom) at $\beta=1.750$.
 Blue, red, and green dots represent the configurations in which $|Q|$
 are ``stable'', ``decreasing'', and ``increasing''.
 The dotted lines represent the Bogomolnyi bound.
 }
 \label{fig:q_vs_action-b1750}
\end{figure}

We studied 
the correlation between the topological charge $Q$ and the value of the action $S_g$.
At the same time, we also investigated the direction of change of $Q$ per
one step of the smearing, by classifying each configuration at a given
$n_{\rm APE}$ into three classes:
\begin{itemize}
 \item ``{\it stable}'' if the change is small, {\it i.e.}
       $|Q(n_{\rm APE})-Q(n_{\rm APE}-1) |\le 0.05$.       
     \end{itemize}
     If not stable, 
    \begin{itemize}
 \item ``{\it decreasing}'' if $Q(n_{\rm APE})$ is approaching zero,
 \item ``{\it increasing}'' if $Q(n_{\rm APE})$ is moving away from zero.
\end{itemize}
Fig.~\ref{fig:q_vs_action-b1750} shows the scatter plot for $Q$ and $S_g$ at several values of $n_{\rm
APE}$, obtained at $\beta=1.750$, where the Bogomolnyi bound,
$S_g=8\,\pi^2|Q|/g^2$ is shown by dotted lines.
The ``stable'', ``decreasing'' and ``increasing'' data points are shown
in blue, red and green, respectively.
There is no qualitative difference in the same plot for other values of 
$\beta$. 
It is seen that points gradually accumulate on integer values of $Q$
by ``increasing'' or ``decreasing''.
The value of the action is never below the Bogomolnyi bound
in each topological sector, as expected.
This indicates that ``increasing'' data can not exist around the
boundary because the smearing lowers the action value and only either
instantons(s) or anti-instanton(s) can exist on the bound.
It is also seen that the larger the value of $|Q|$ is, faster the
minimum of the action in the topological sector reaches the
Bogomolnyi bound.
Thus, the minimum value of $S_g$ in the $Q=0$ sector arrives at the
bound, {\it i.e.} $S_g=0$, last.

Instantons are known to saturate the Bogomolnyi bound.
Therefore, the data points with nonzero $Q$ around the bound are
attributed to approximate instantons or anti-instantons, and the ``decreasing''
occurring around there are interpreted as (anti-) instantons ``falling
through the lattice''.
We expect that all the ``increasing'' and ``decreasing'' in the second
phase are caused by ``falling''.
In order to examine this expectation, we introduce the participation
ratio defined by
\begin{align}
   P(n_{\rm APE})
&:= \frac{1}{N_{\rm site}}
   \frac{\displaystyle
         \left(\sum_x q(x,n_{\rm APE})^2\right)^2}
        {\displaystyle \sum_x q(x,n_{\rm APE})^4} \ ,
\end{align}
where $q(x,n_{\rm APE})$ denotes the topological charge density $q(x)$
in eq.~(\ref{eq:qdensity}) after $n_{\rm APE}$ steps of smearing.
The participation ratio takes a value between $1/N_{\rm site}$ and 1.
The maximal value $P(n_{\rm APE})=1$ is realized when $q(x,n_{\rm APE})$ takes a flat
distribution over the whole space-time.
On the other hand, the possible minimum value, $1/N_{\rm site}$, is
attained when the density forms a local peak,
$q(x,n_{\rm APE})=\delta(x)$. 
\begin{figure}[b]
 \begin{center}
 \vspace{-5ex}
  \begin{tabular}{c}
  \includegraphics[width=0.8 \textwidth]
  {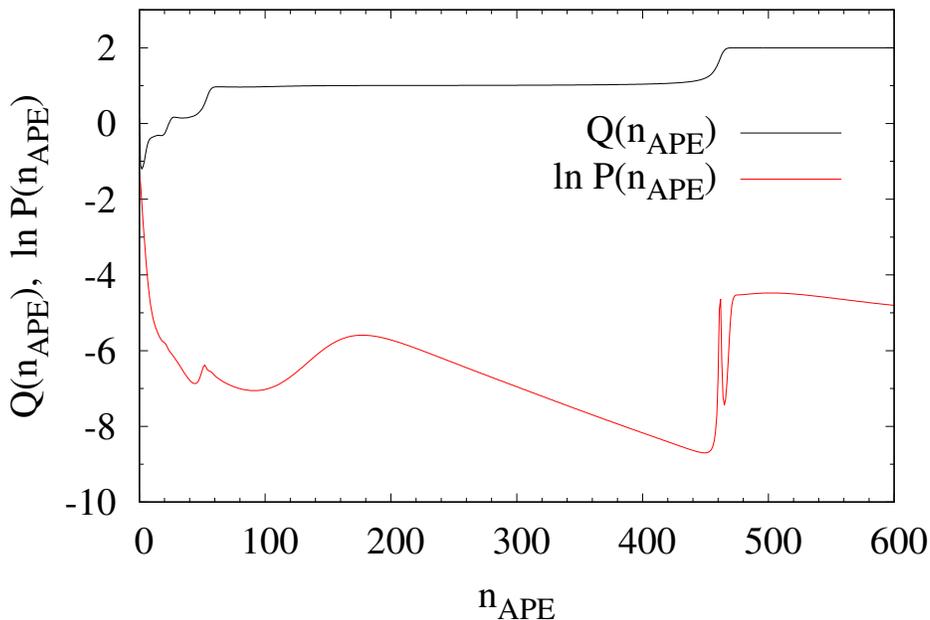}
  \end{tabular}
 \vspace{5ex}
 \end{center}
 \caption{
 An example of $Q(n_{\rm APE})$ and $\ln P(n_{\rm APE})$ as a function
 of $n_{\rm APE}$ at $\beta=1.850$.
 }
 \label{fig:participation-ratio}
\end{figure}
Fig.~\ref{fig:participation-ratio} shows the smearing history of $Q$
and $\ln P$ as a function of $n_{\rm APE}$ for one particular
configuration at $\beta=1.850$.
For $n_{\rm APE}\simge 30$, whenever $Q$ changes, $\ln P$ shows a rapid
increase after slow decrease over many smearing steps.
This can be interpreted as that a local object in topological charge
density gradually shrinks and suddenly disappears at some point with a
change of $Q$.
This is precisely what happens when the ``falling through the lattice''
occurs~\cite{BilsonThompson:2004ez}.

\begin{figure}[tbp]
 \begin{center}
   \vspace*{-0ex}
  \begin{tabular}{cc}
   \hspace*{-2ex}
  \includegraphics[width=0.5 \textwidth]
  {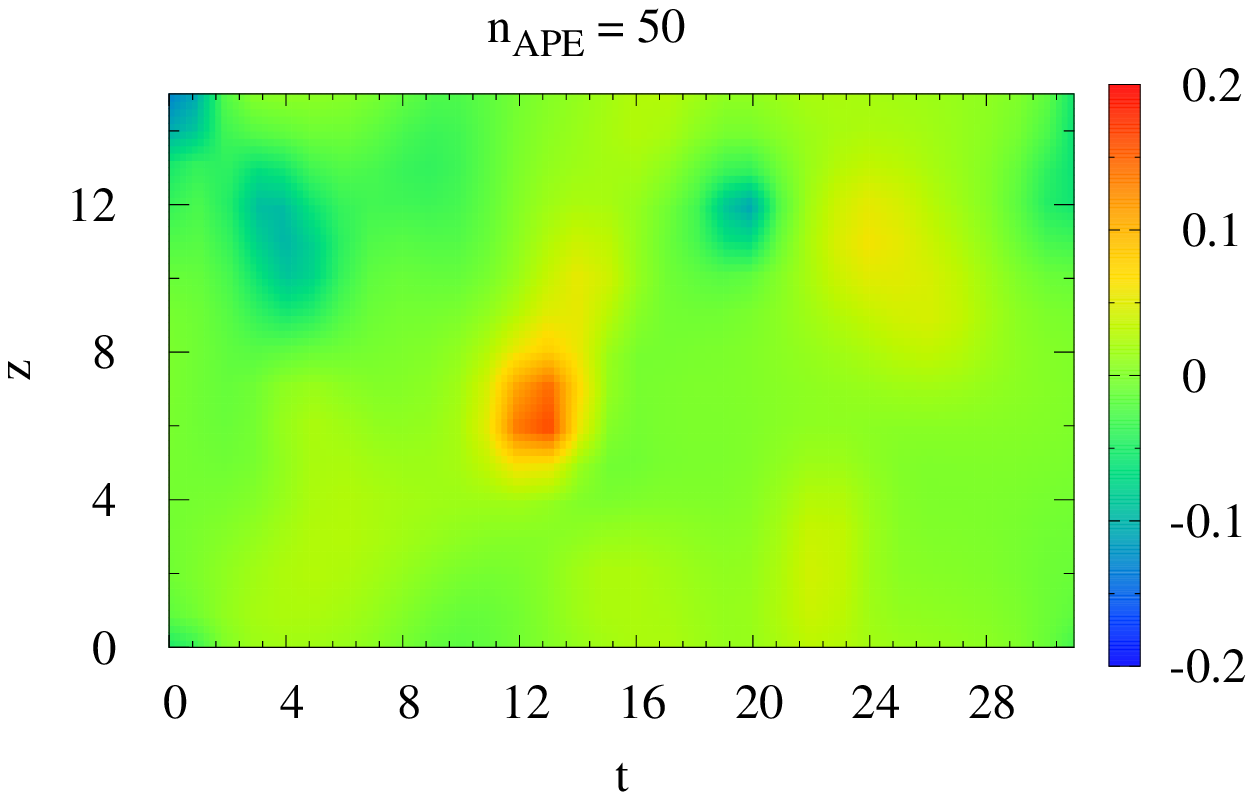}&
   \hspace*{-2ex}
  \includegraphics[width=0.5 \textwidth]
  {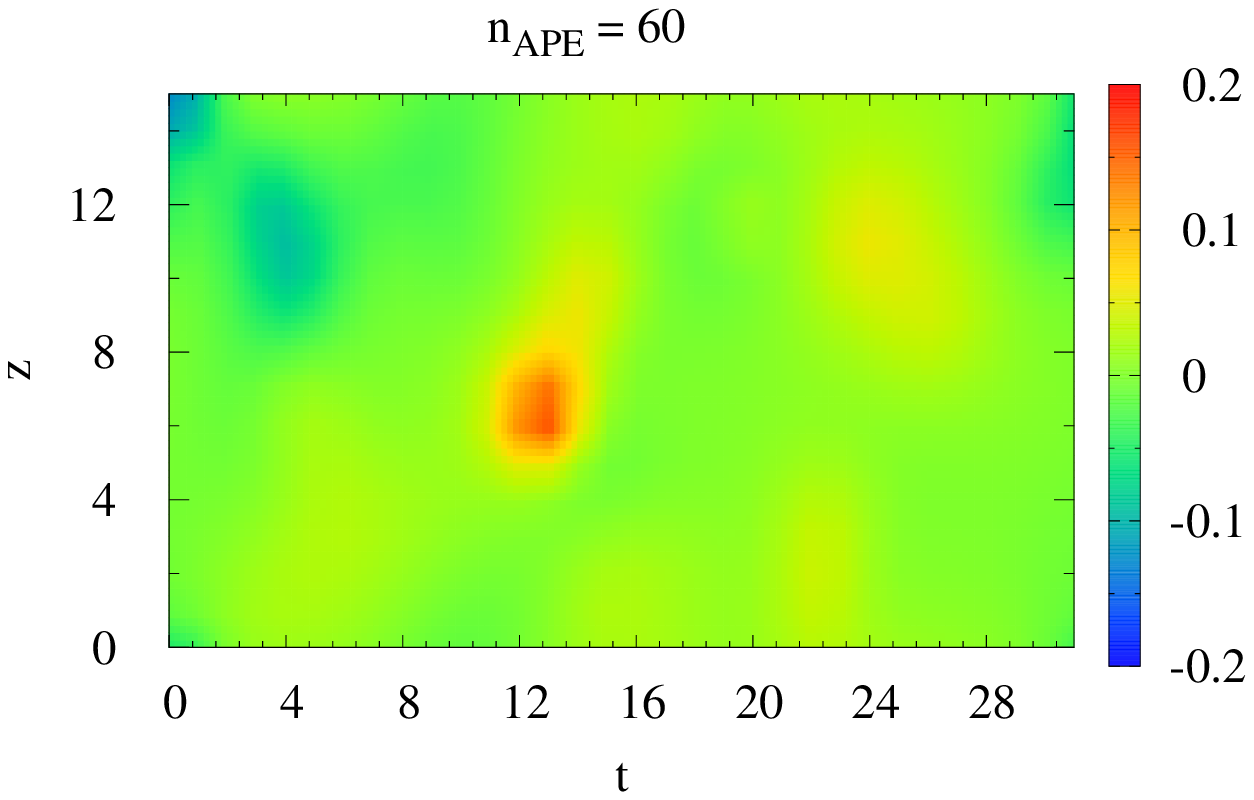}
   \vspace*{-2ex}\\
   \hspace*{-2ex}
  \includegraphics[width=0.5 \textwidth]
  {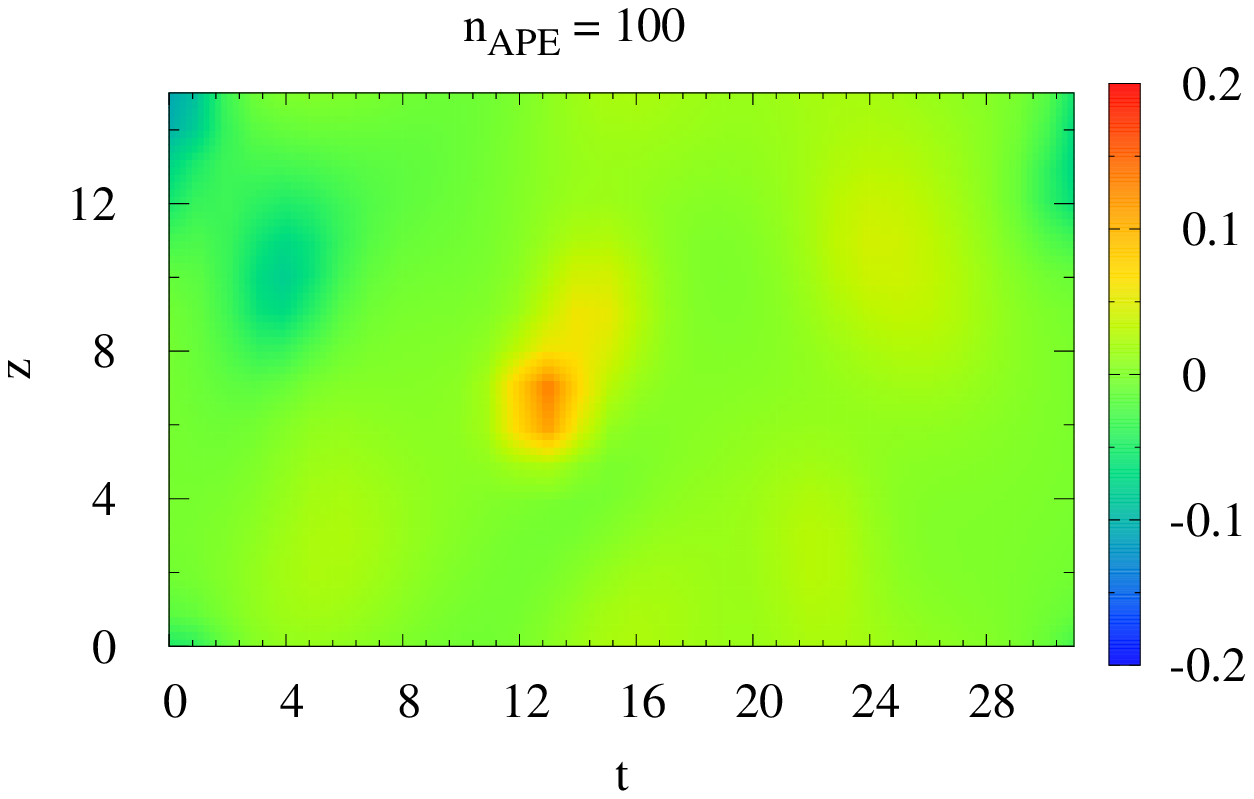}&
   \hspace*{-2ex}
  \includegraphics[width=0.5 \textwidth]
  {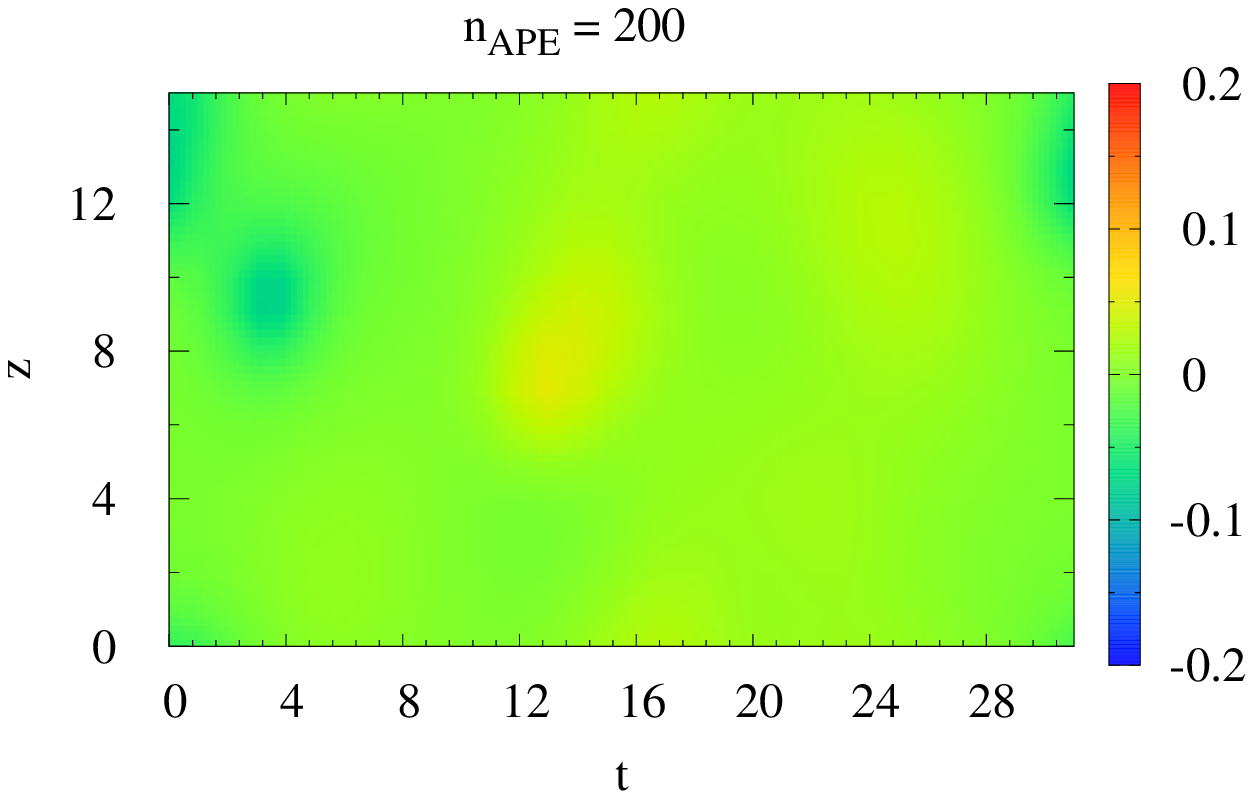}
   \vspace*{-2ex}\\
   \hspace*{-2ex}
  \includegraphics[width=0.5 \textwidth]
  {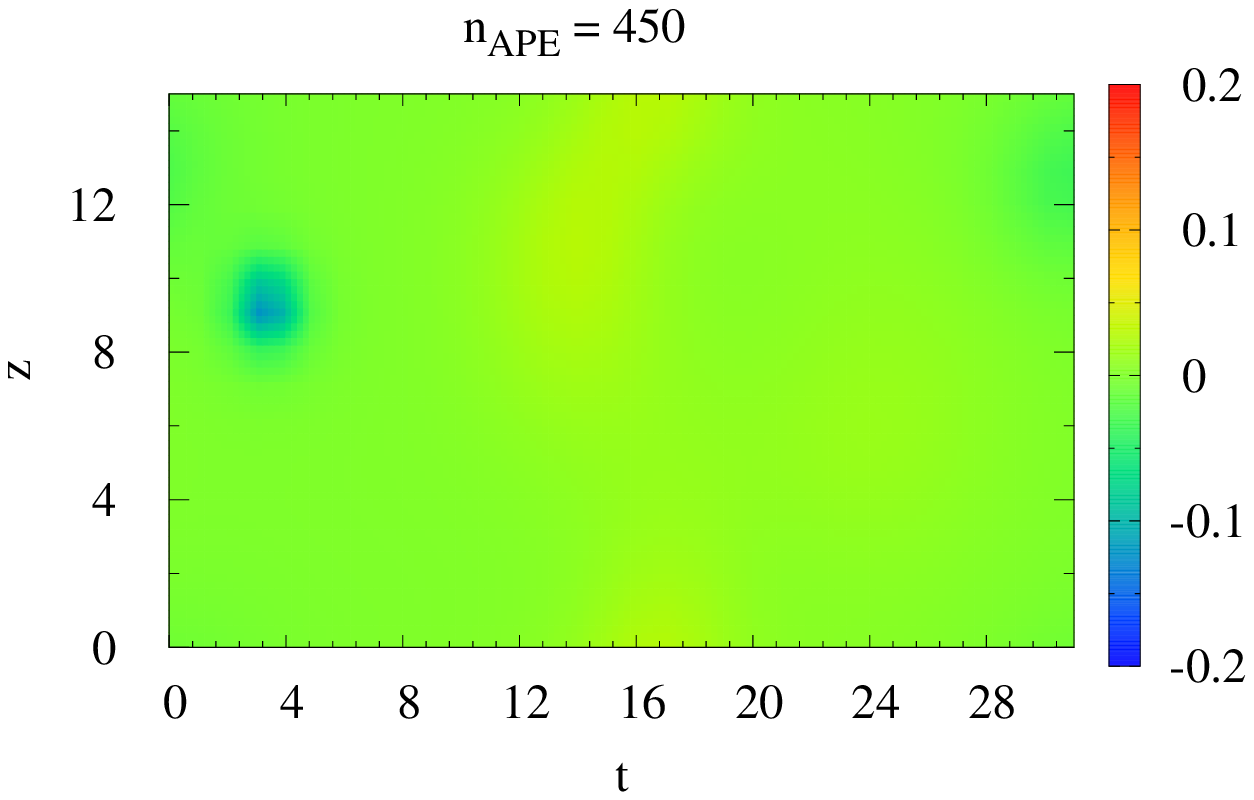}&
   \hspace*{-2ex}
  \includegraphics[width=0.5 \textwidth]
  {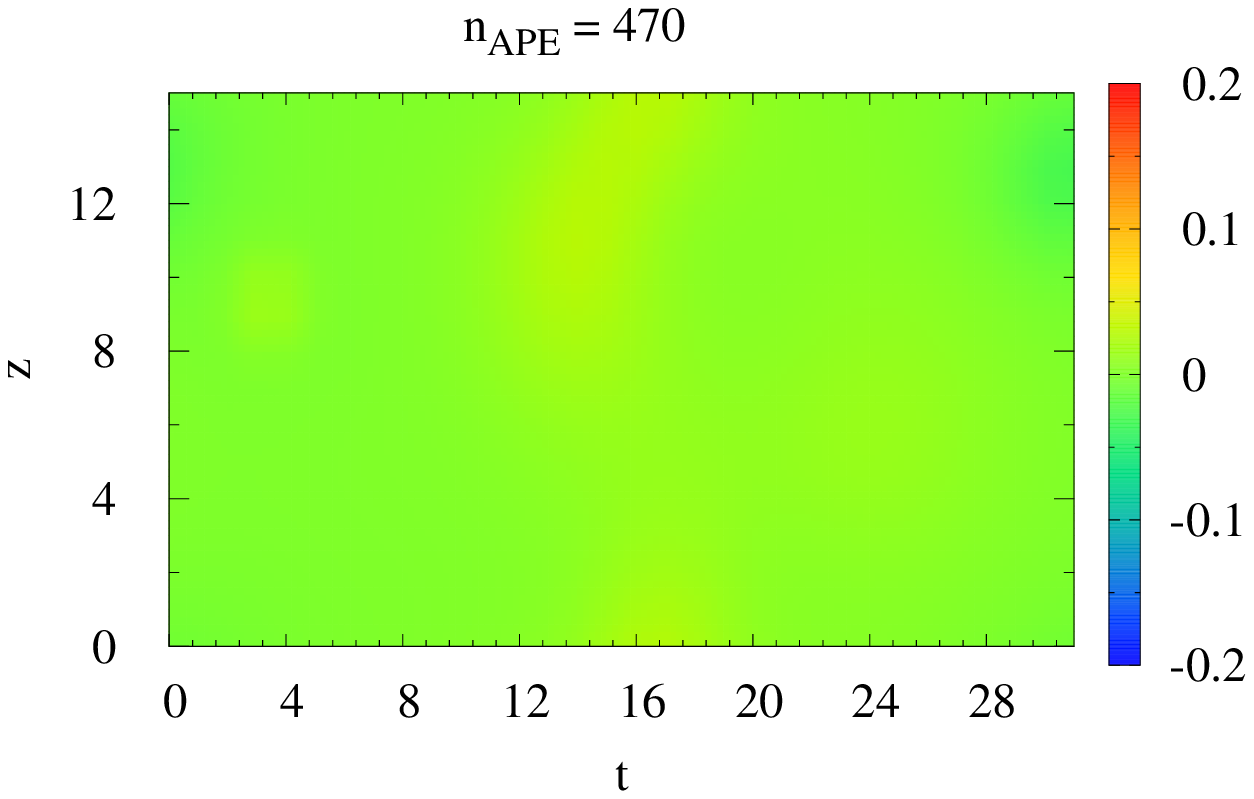}
  \end{tabular}  
   \vspace{2ex}
 \end{center}
 \caption{Distribution of topological charge projected onto $z$-$t$
 plane at $n_{\rm APE}=50$, 60, 100, 200, 450, and 470.
 }
 \label{fig:q-distribution}
\end{figure}

We can directly check this interpretation by studying the distribution of the topological charges.
Fig.~\ref{fig:q-distribution} shows the topological charge density,
projected onto the $z$-$t$ plane, of the same configuration as in
Fig.~\ref{fig:participation-ratio}.
Between $n_{\rm APE}=50$ and 60 and $n_{\rm APE}=450$ and 470, $Q$
increases by unity, at the same time a negative peak disappears.
Between $n_{\rm APR}=100$ and 200, a positive peak seems to be smeared
but does not suddenly disappear.
It seems that a complicated process such as a pair annihilation happens in
the latter case.

From these observations, we conclude that the changes of $Q$ occurring in
the second phase are dominated by the ``falling'' of instantons or
anti-instantons. We expect that the ``falling'' occurs also in the first phase, but it is overshadowed
by changes originating from other reasons.

\begin{figure}[tbhp]
 \begin{center}
 \vspace{-5ex}
  \begin{tabular}{c}
  \includegraphics[width=0.8 \textwidth]
  {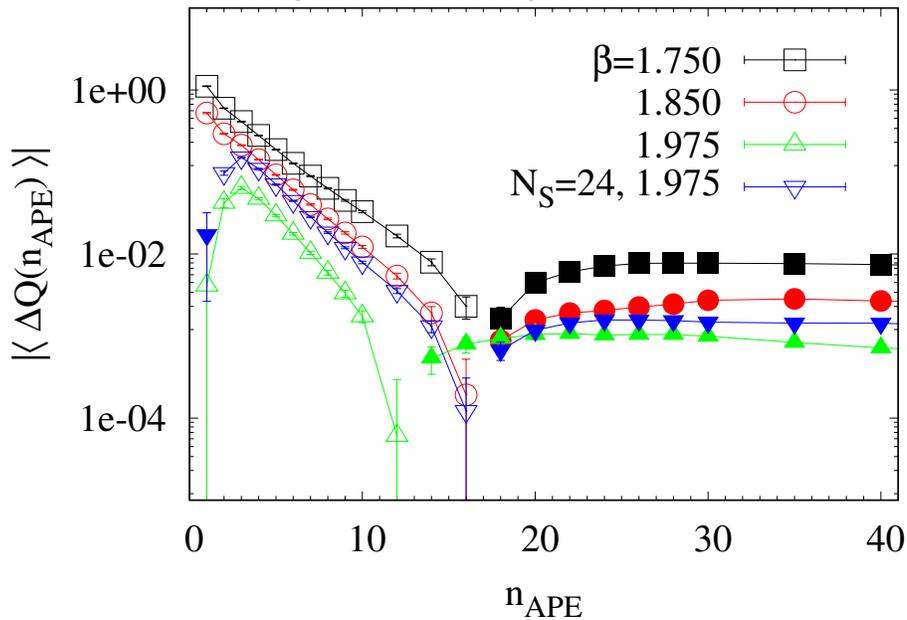}\\
  \end{tabular}  
 \vspace{5ex}
 \end{center}
 \caption{The {\it decay rate} of the topological charge.
 }
 \label{fig:dqdnape-lat}
\end{figure}

Instanton and anti-instantons will ``fall'' at an equal rate.
In configurations with $Q>0$, more instantons exist than anti-instantons
and vice versa.
Then, it is expected that the ``decreasing'' would happen more frequently
than the ``increasing'' in the second phase.
To see if this is the case, we calculate the ensemble average of
\begin{align}
 \Delta Q(n_{\rm APE}) =\left\{\begin{array}{ll}
Q(n_{\rm APE}+1)-Q(n_{\rm APE}) & \mbox{ for }	Q(n_{\rm APE})>0\\
Q(n_{\rm APE})-Q(n_{\rm APE}+1) & \mbox{ for }  Q(n_{\rm APE})<0
		  \end{array}
\right.\ .
\end{align}
The sign of $\Delta Q(n_{\rm APE})$ tells us which of ``increasing'' or
``decreasing'' happens when going from $n_{\rm APE}$ to $n_{\rm APE}+1$.
Fig.~\ref{fig:dqdnape-lat} shows the $n_{\rm APE}$ dependence of
$|\langle\Delta Q(n_{\rm APE})\rangle|$, where the symbols are filled
when its original value is negative.
The results from four ensembles show exponential fall with approximately
a common exponent for $n_{\rm APE}\simle 10$, while they take almost
constant negative values for $n_{\rm APE}\simge 20$.
The result for $\beta=1.975$ and $N_S=16$ (triangle-up) shows slightly
different behaviors probably because of the small physical volume.
At any rate, this plot clearly shows that the boundary separating the
two phases is located $n_{\rm APE}\sim 20$.
In the following analysis, we only deal with the data for
$n_{\rm APE} \ge 20$, where the short distance fluctuations are gone.

Before closing this subsection, let us add one comment.
In Ref.~\cite{Horvath:2003yj}, the shape of topological objects in
SU(3) gauge theory is examined, and low dimensional long range structures
rather than local lumps are discovered.
Note that the analysis presented above does not indicate anything about
the shape because the smearing changes it.
Clearly, it is interesting to perform a similar study in the SU(2)
case because the analysis performed in
Refs.~\cite{Ahmad:2005dr,Lian:2006ky} suggests that the structure could
be more localized for SU(2) than for SU(3).

\subsection{Results}
\label{subsec:results}

\begin{figure}[b]
 \begin{center}
\vspace*{-2ex}
  \begin{tabular}{cc}
  \includegraphics[width=0.45 \textwidth]
  {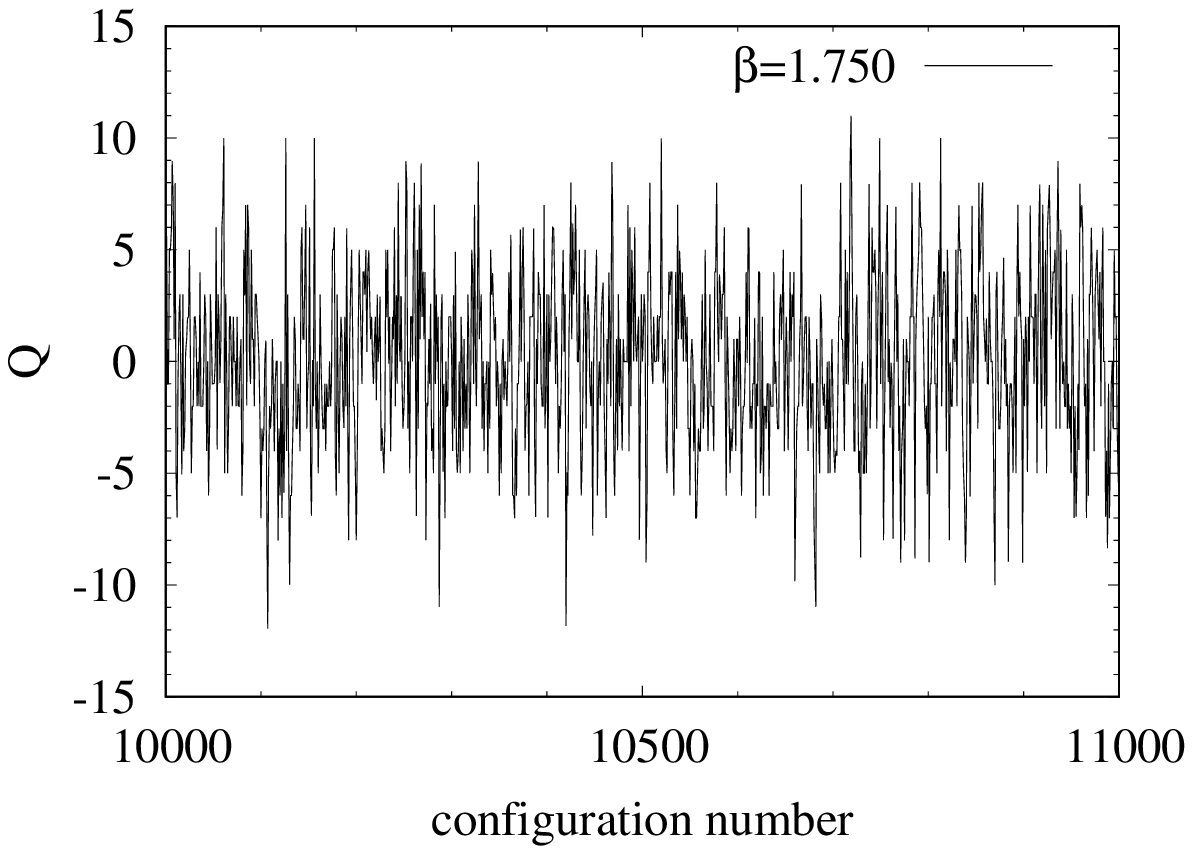} &
  \includegraphics[width=0.45 \textwidth]
  {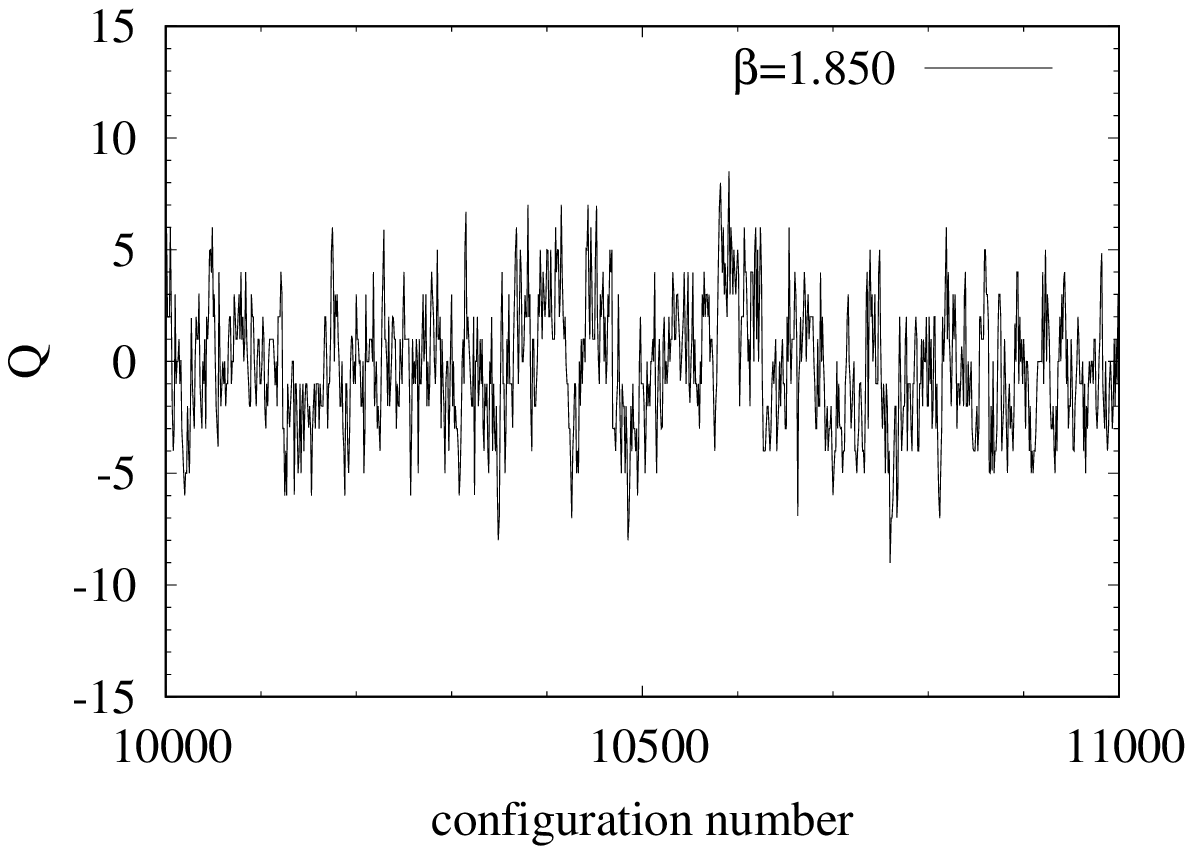} \\
  \includegraphics[width=0.45 \textwidth]
  {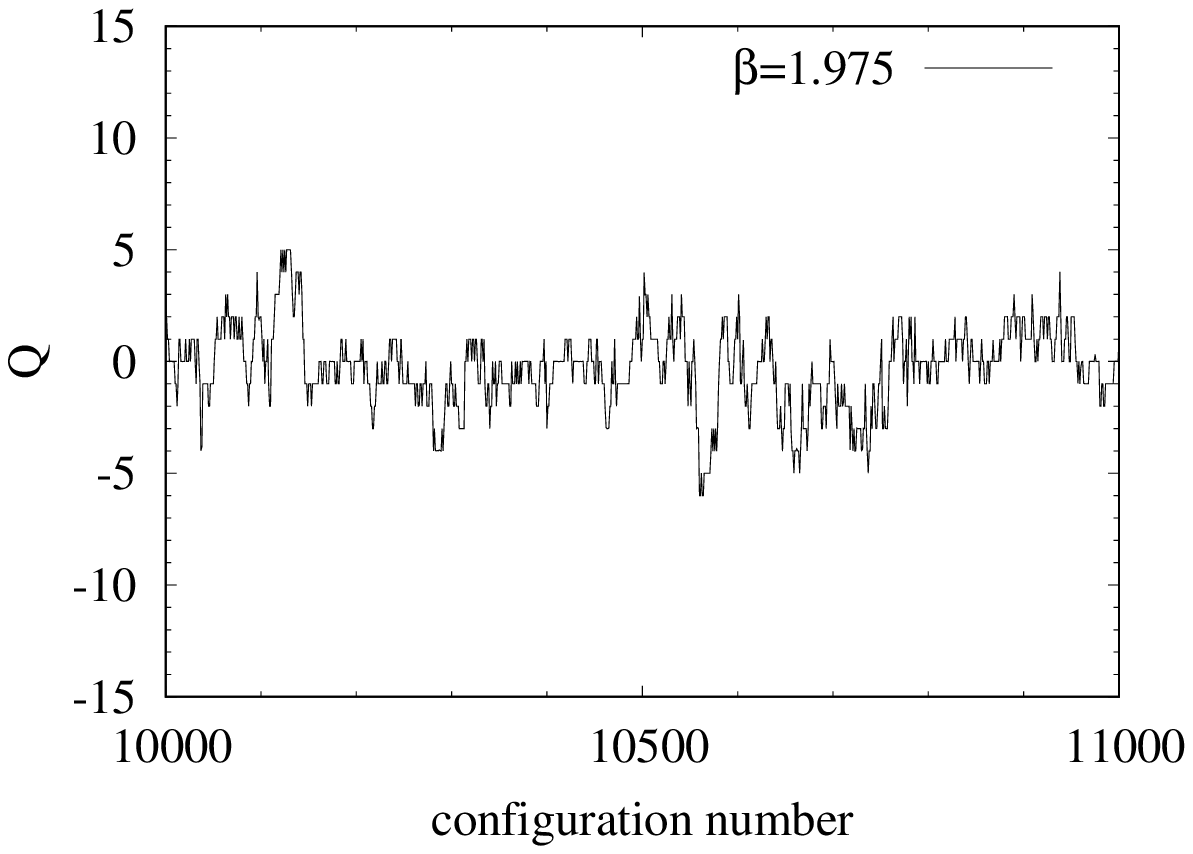} &
  \includegraphics[width=0.45 \textwidth]
  {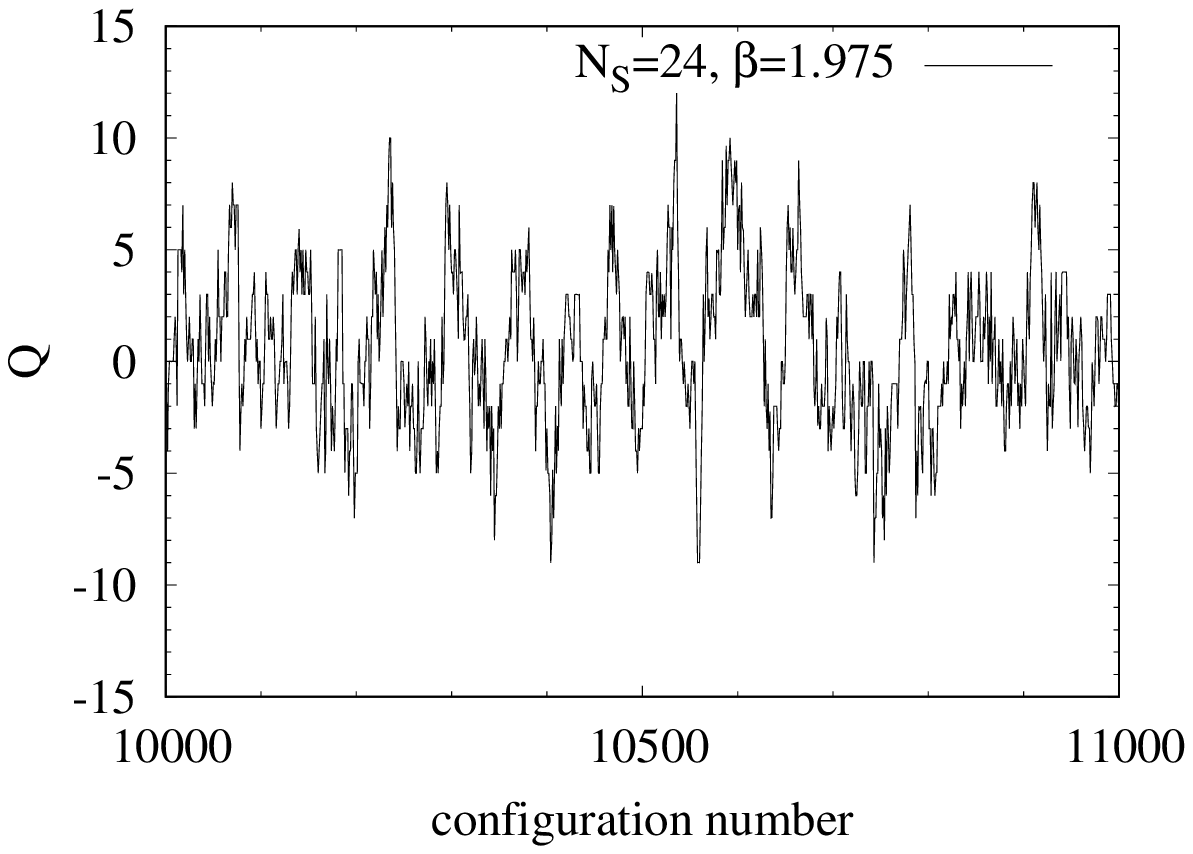} \\
  \end{tabular}  
\vspace{2ex}
 \end{center}
 \caption{Monte Carlo history of $Q$ at four ensembles.
 }
 \label{fig:Q-history}
\end{figure}

Fig.~\ref{fig:Q-history} shows the Monte Carlo history of $Q$ over
a  thousand configurations in four ensembles obtained at $n_{\rm APE}=800$.
It is seen that the fluctuation of $Q$ is frequent enough, and that the
amplitude depends on $\beta$ and $N_{\rm site}$.
In the following analysis, all the measurements are binned with the bin
size of 100 configurations, and a single elimination jackknife method is
used to estimate statistical uncertainties.

\begin{figure}[htbp]
\vspace{-3ex}
 \begin{center}
  \begin{tabular}{cc}
  \includegraphics[width=0.45 \textwidth]
  {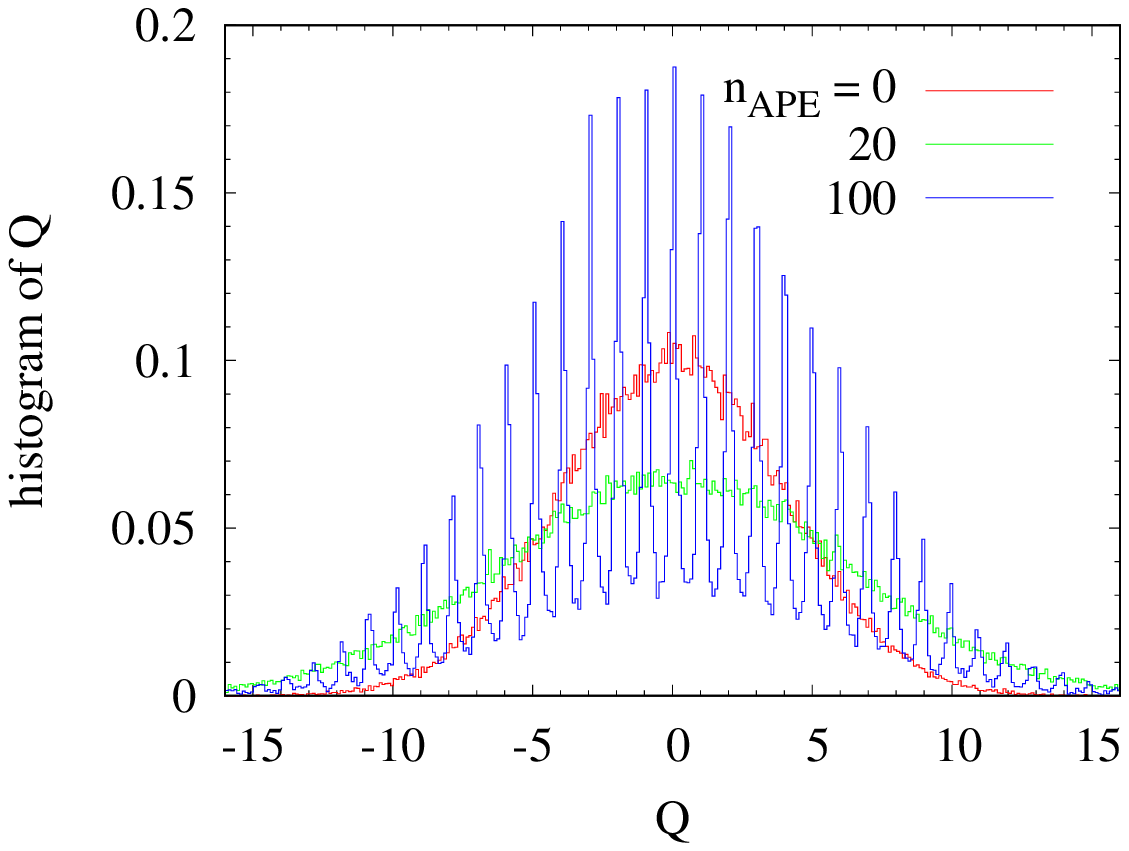} &
  \includegraphics[width=0.45 \textwidth]
  {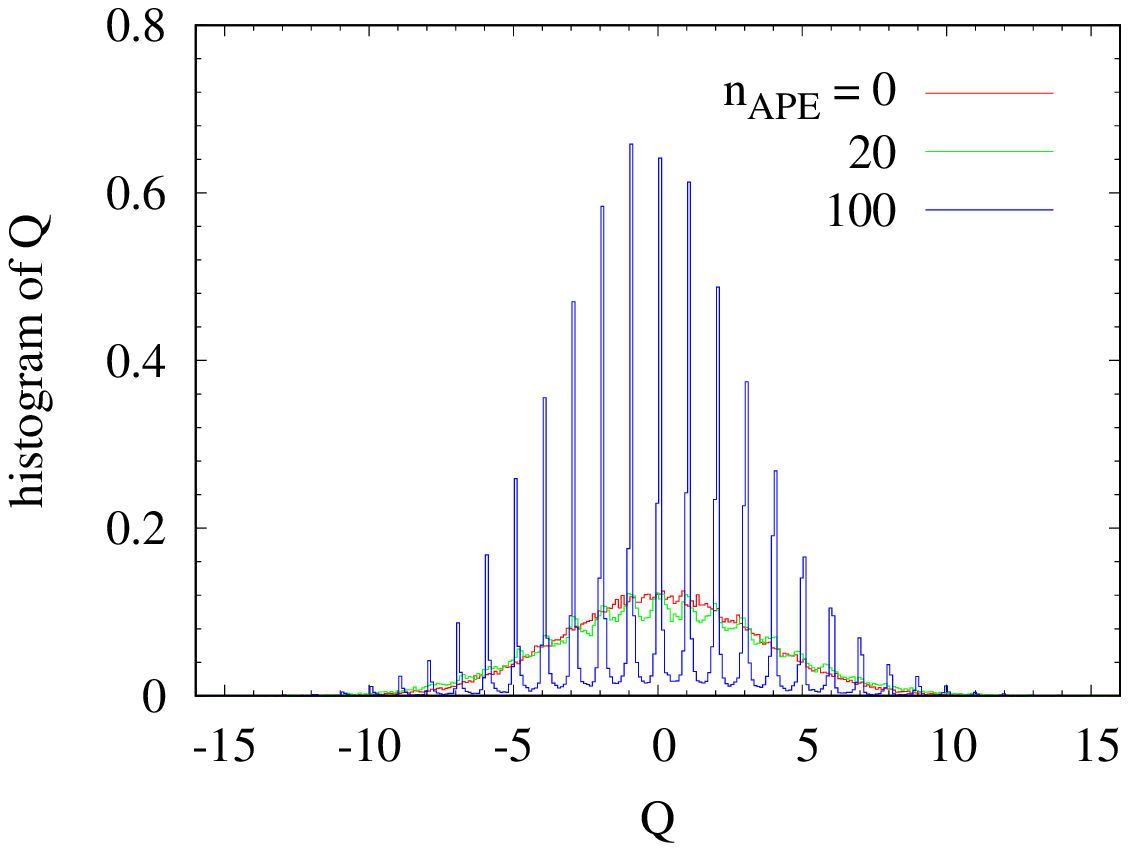} \\
  \includegraphics[width=0.45 \textwidth]
  {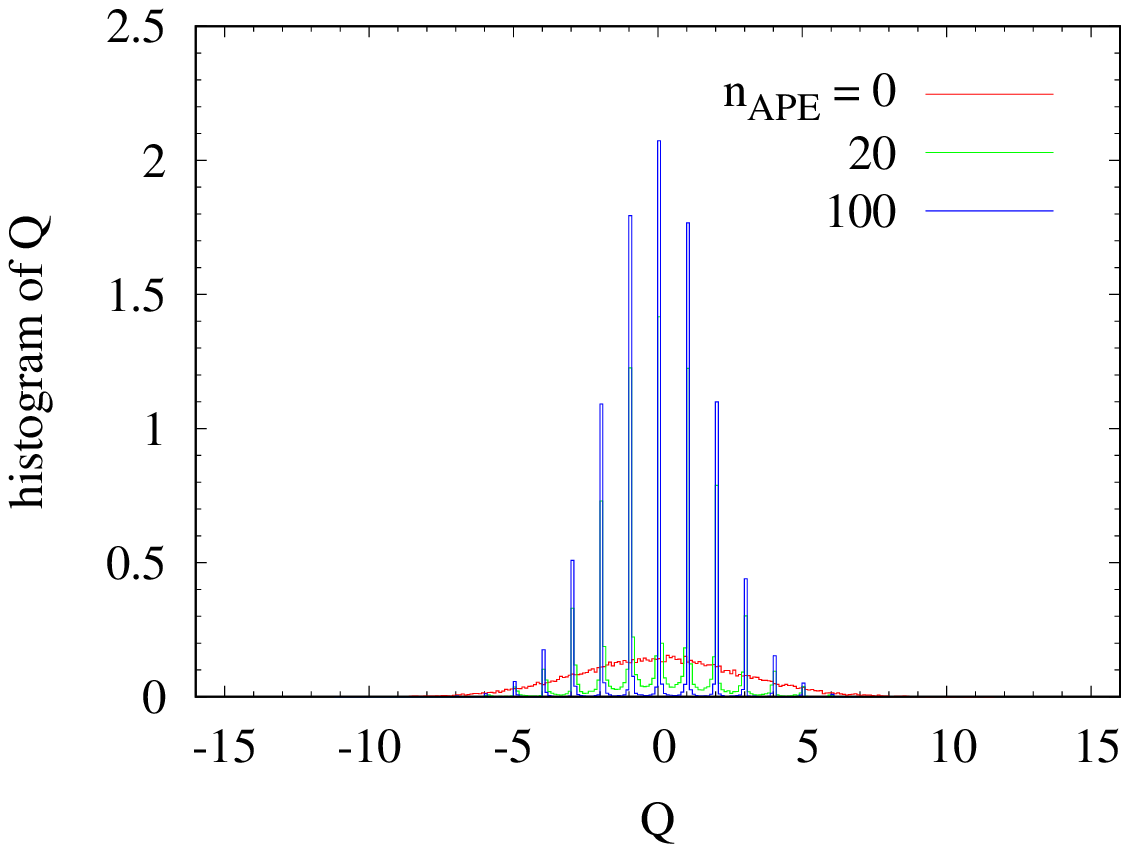} &
  \includegraphics[width=0.45 \textwidth]
  {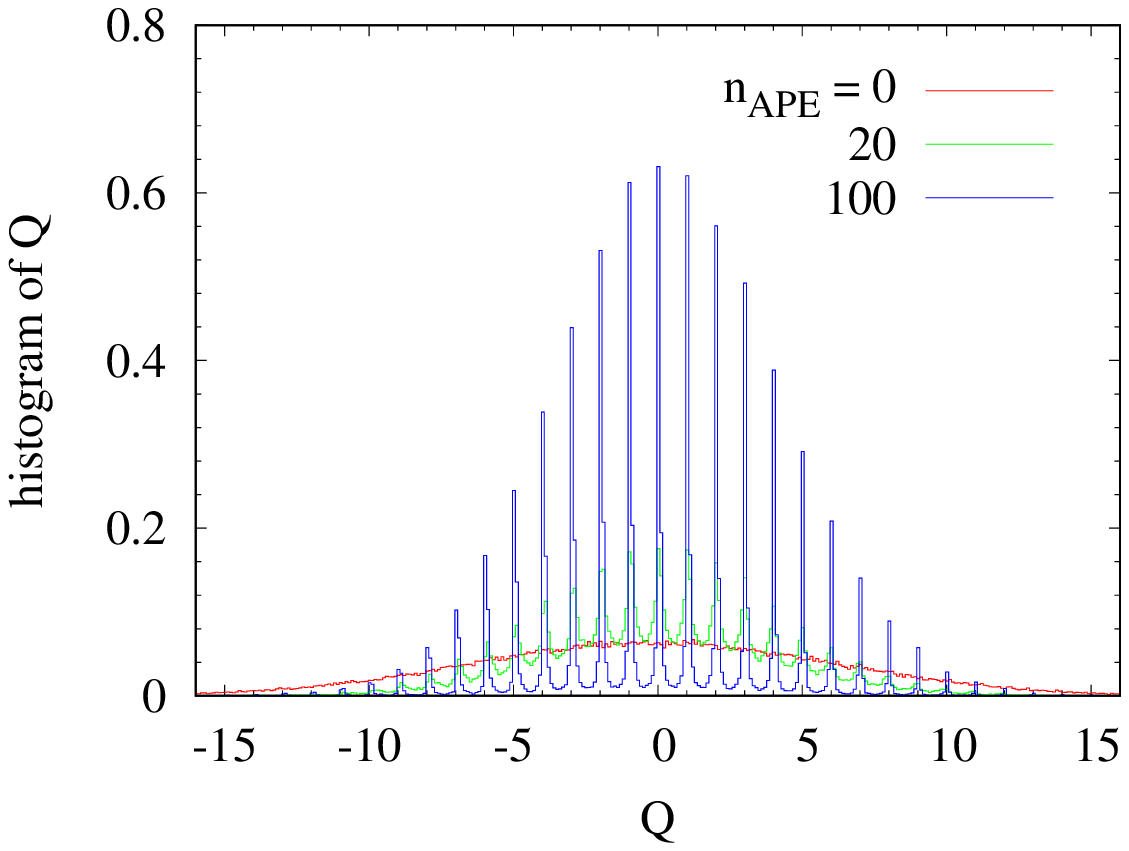} \\
  \end{tabular}  
\vspace{3ex}
 \end{center}
 \caption{Histogram of $Q$ for four ensembles at $n_{\rm APE}=0$, 20,
 100.
 }
 \label{fig:Q-histogram}
\end{figure}

Fig.~\ref{fig:Q-histogram} shows the histogram of $Q$ for four
ensembles at $n_{\rm APE}=0$, 20, 100.
Approximate Gaussian shape is seen in all ensembles.

\begin{figure}[htbp]
 \begin{center}
\vspace{-3ex}
  \begin{tabular}{c}
  \includegraphics[width=0.7 \textwidth]
  {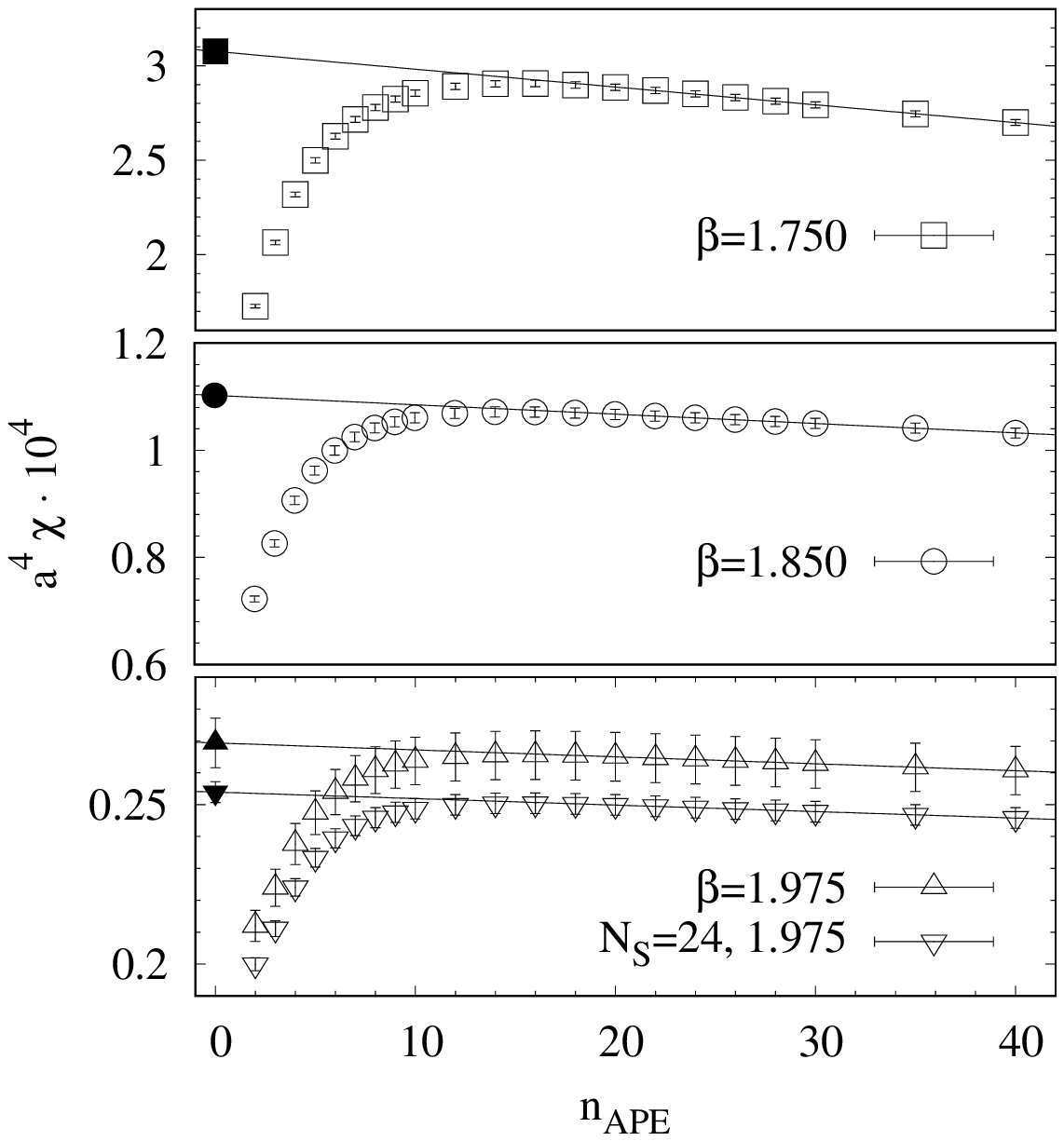}
\vspace{3ex}
  \end{tabular}  
 \end{center}
 \caption{$a^4 \chi$ for the four ensembles as a function of
 $n_{\rm APE}$.
 }
 \label{fig:Q2}
\end{figure}

Fig.~\ref{fig:Q2} shows the topological susceptibility in lattice
unit, $a^4\chi(n_{\rm APE})=\langle Q^2 \rangle/N_{\rm site}$, as a
function of $n_{\rm APE}$.
A mild decrease is seen for $n_{\rm APE}\ge 20$ as expected from a
negative constant observed in Fig.~\ref{fig:dqdnape-lat}.
We determine topological susceptibility at each lattice by extrapolating
the smeared data in the second phase to $n_{\rm APE}\to 0$ because the
``falling'' is supposed to take place even in the first phase.
The data points in $n_{\rm APE}\in[20,40]$ are well described by a linear
function,
\begin{align}
 a^4 \chi(n_{\rm APE}) = a^4 \chi(0) + c_1\, n_{\rm APE}\ .
\end{align}
The fit results are tabulated in Tab.~\ref{tab:Q2ave-fit}.
\begin{figure}[ht]
 \begin{center}
\vspace{-3ex}
  \begin{tabular}{c}
  \includegraphics[width=0.7 \textwidth]
  {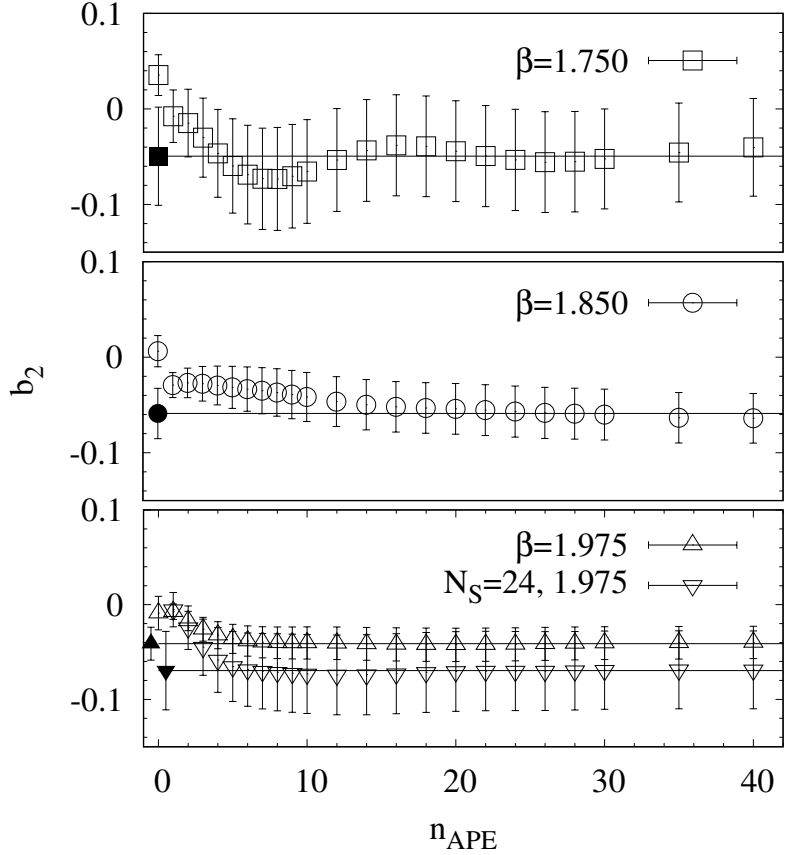}
  \end{tabular}  
\vspace{3ex}
 \end{center}
 \caption{$n_{\rm APE}$ dependence of $b_2$.
 }
 \label{fig:b2}
\end{figure}
\begin{table}[b]
 \begin{tabular}{p{8ex}p{6ex}|ccc}
  \,\,\, $\beta$ & $N_S$ &\ $a^4 \chi(0)\times 10^4$\ &
  \ $c_1\times 10^7$\ &\ $b_2(0)\times 10^2$\ \\
  \hline
  1.750 & 16 & $3.08(2)$ & $-9.4(3)$   & $-5(5)$ \\
  1.850 & 16 & $1.10(1)$ & $-1.8(1)$   & $-6(3)$ \\
  1.975 & 16 & $0.269(8)$ & $-0.22(2)$ & $-4(2)$ \\
  1.975 & 24 & $0.254(3)$ & $-0.20(1)$ & $-7(4)$ \\
 \end{tabular}
 \caption{Fit results.}
 \label{tab:Q2ave-fit}
\end{table}

Fig.~\ref{fig:b2} shows $n_{\rm APE}$ dependence of $b_2$.
Since $b_2$ is found to be constant for $n_{\rm APE}\ge 20$, we perform
the constant fit to extract $b_2$ at $n_{\rm APE}=0$.
The fit results are shown in Tab.~\ref{tab:Q2ave-fit}.

The values of $b_2$ obtained at $\beta=1.975$ with two lattice volumes turns out to be
consistent with each other due to the large statistical uncertainty,
while 1.8 $\sigma$ difference is observed for $\chi$.
In Ref.~\cite{DelDebbio:2002xa,Bonati:2016tvi}, these quantities are
calculated with several different volumes for SU($N$) with $N=3$, 4, 6
down to $L\,\sigma_{\rm str}\sim 2.5$, and no finite volume effect is
observed.
Our lattice with $\beta=1.975$ and $N_S=16$ corresponds to
$L\,\sigma_{\rm str} = 2.4$ (see Tab.~\ref{tab:parameters}), which is
smaller than but close to 2.5 and hence finite volume effects, if any,
should not be significant.
Thus, 1.8 $\sigma$ difference observed at $\beta=1.975$ is considered as
a statistical fluctuation, and we include both results in the following
analysis.

\begin{figure}[tbp]
 \begin{center}
  \begin{tabular}{cc}
  \includegraphics[width=0.5 \textwidth]
  {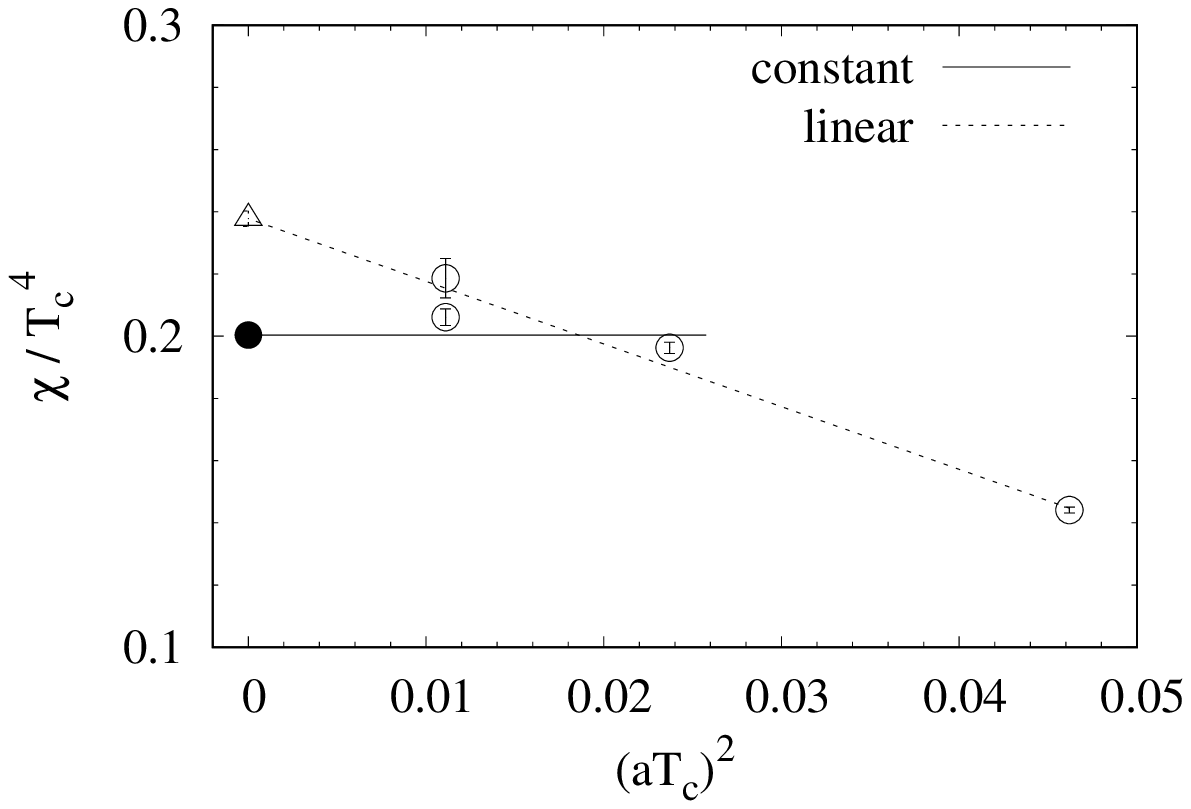} &
  \includegraphics[width=0.5 \textwidth]
  {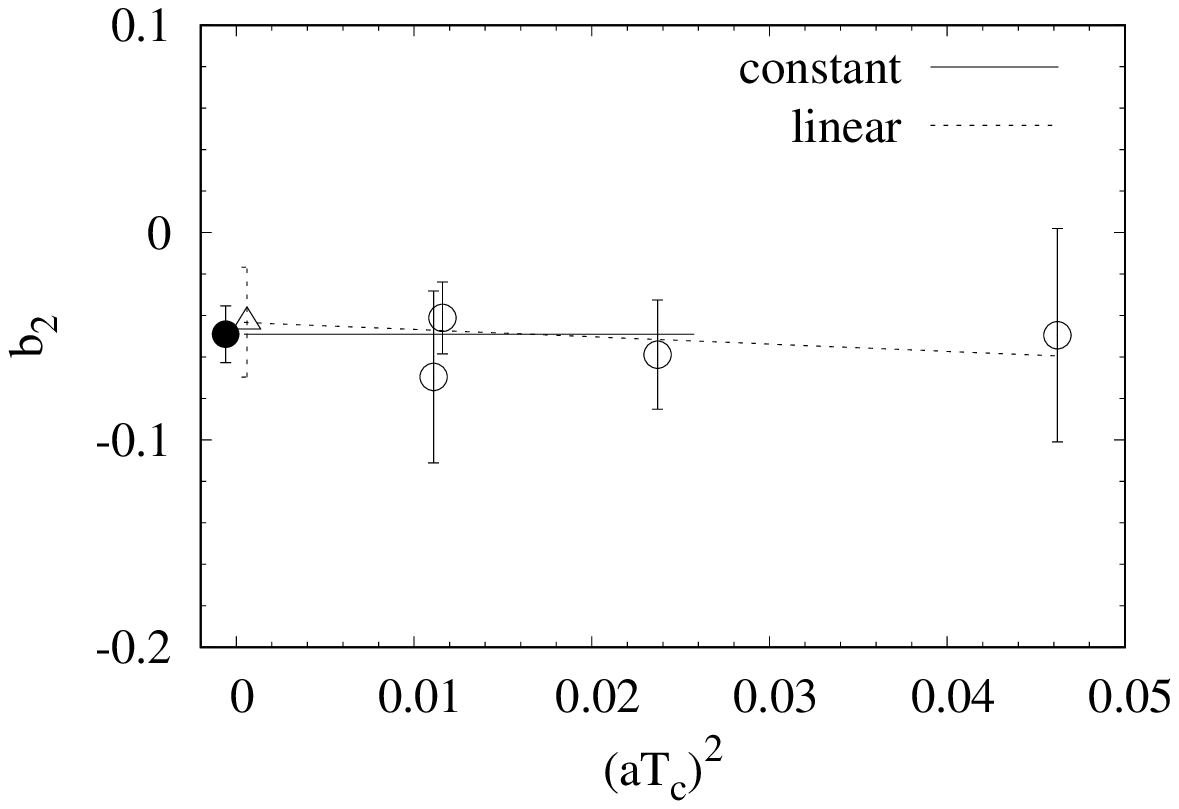} \\
  \end{tabular}  
 \end{center}
 \caption{The continuum limit of $\chi/T_c^4$ and $b_2$.
 The solid lines in both plots are the results from a constant fit using
 only two finer lattices, and the dashed lines are those from a linear
 fit using all lattices.
 }
 \label{fig:cont-limit}
\end{figure}

Next we discuss the continuum limit.
Fig.~\ref{fig:cont-limit} shows the extrapolation of $\chi/T_c^4$ and
$b_2$ to the continuum.
The limit for both quantities is examined by applying two
functional forms:
\begin{enumerate}
 \item constant excluding the coarsest lattice
 \item linear in $a^2$ using all lattices .
\end{enumerate}

The constant fit is used to estimates the continuum limit assuming no
scaling violation for the finer lattices, whereas the linear fit
including the coarsest lattice serves to probe the possible size of the
scaling violation.
Indeed, it turns out that these two extrapolations yield smallest and
largest values for $\chi/T_c^4$ among other reasonable choices and thus
the difference provides with the conservative estimate for the scaling
violation.

As the continuum limit of $\chi/T_c^4$, we obtain $0.200(1)$ and
$0.238(3)$ for the constant and linear fit, respectively.
The difference is due to the large deviation at the coarsest lattice
from those at finer lattices, and is interpreted as the potential size
of the scaling violation.
As for $b_2$, the constant and linear fits lead to $-0.049(14)$ and
$-0.043(27)$, respectively, and the statistical error turns out to
dominate the systematic one.

The final results thus obtained are
\begin{align}
&\frac{\chi}{T_c^4} = 0.200(39)\ ,\qquad
\frac{\chi^{1/4}}{T_c} = 0.674(31)\ ,\qquad
b_2= -0.049(20)\ ,
\label{eq:our-result}
\end{align}
where the constant fit is used to estimate the central value and the
statistical and the systematic errors are summed in linear.

Further conservative error estimate may be possible by taking the
statistical error of the linear fit for $b_2$ as the final uncertainty.
On the other hand, the statistical uncertainty is, in general, expected
to increase with complexity of functional form, which could result in
overestimate of uncertainty.
Since $b_2$ behaves as a constant, we here adopt the constant fit to
provide the representative statistical uncertainty.

\begin{figure}[tbp]
 \begin{center}
\vspace{-3ex}
  \begin{tabular}{c}
  \includegraphics[width=0.7 \textwidth]
  {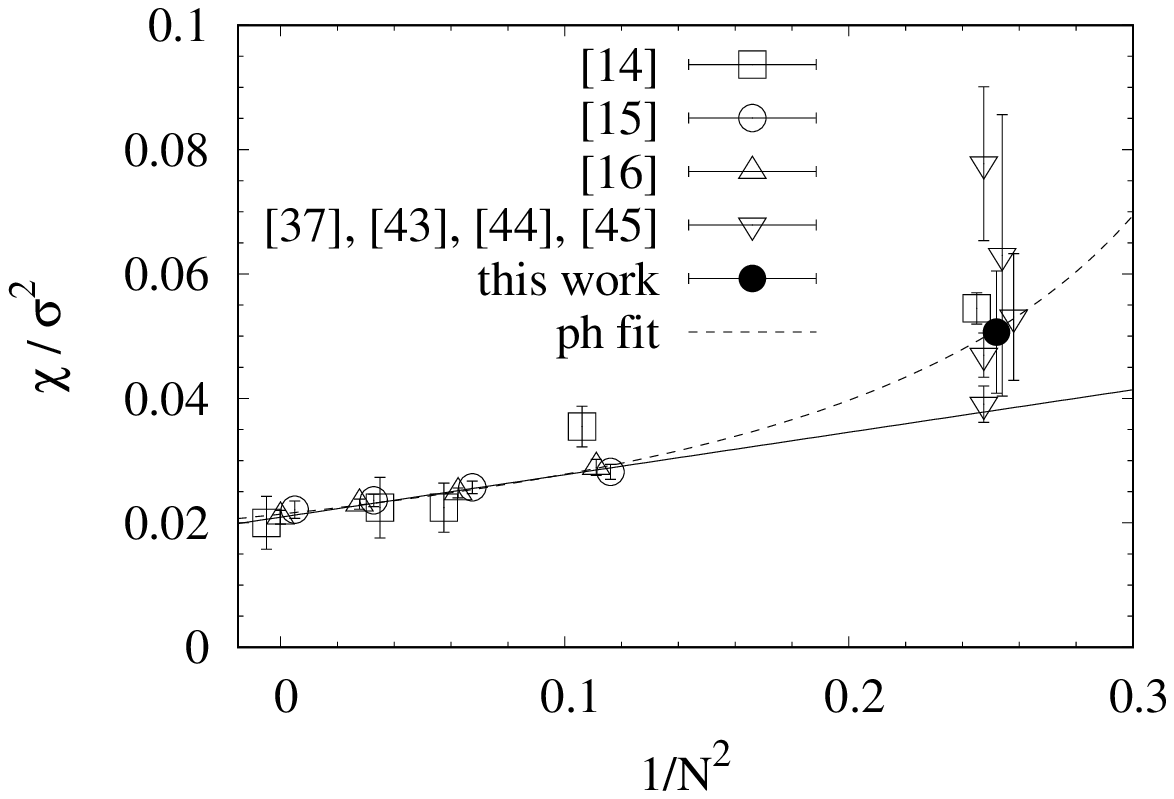} \\
  \includegraphics[width=0.7 \textwidth]
  {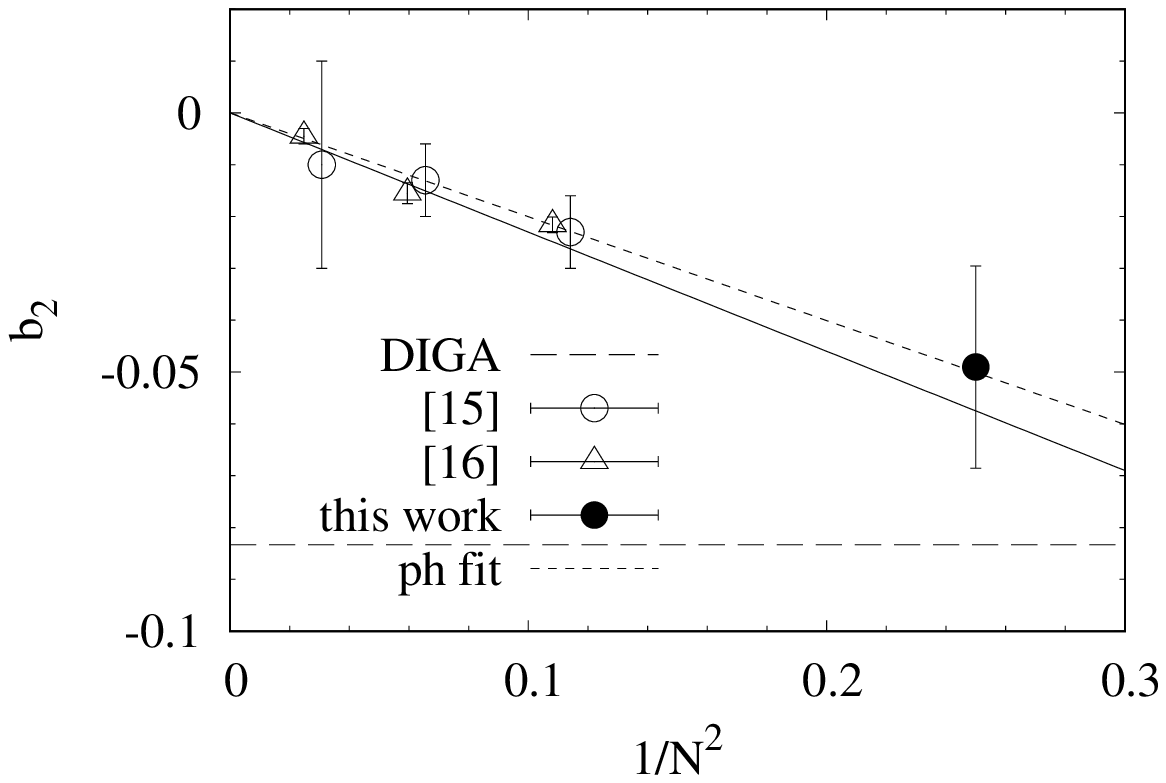}\\
\vspace{3ex}
  \end{tabular}  
 \end{center}
 \caption{The $N$ dependence of $\chi/\sigma_{\rm str}^2$ and $b_2$,
 where the numbers in the legend denote the reference numbers.
 Each data point is slightly shifted horizontally to make it easier to
 see.
 The horizontal dashed line in the $b_2$ plot represents the dilute
 instanton gas approximation (DIGA).}
 \label{fig:ncdep}
\end{figure}

In Refs.~\cite{Lucini:2001ej,DelDebbio:2002xa,Bonati:2016tvi}, the
topological susceptibility $\chi$ is calculated in SU($N$) gauge theory
with several values of $N$ to study the large $N$ behavior.
In Refs.~\cite{deForcrand:1997esx,Alles:1997qe,DeGrand:1997gu,Lucini:2001ej,Berg:2017tqu},
$\chi$ is estimated for SU(2) gauge theory.
As for $b_2$, the $N$ dependence is studied for $N\ge 3$ in
Refs.~\cite{DelDebbio:2002xa,Bonati:2016tvi}.
No result is available in the continuum limit for $N=2$.\footnote{See,
for an exploratory study, Ref.~\cite{Bhanot:1983sn}.} 
Fig.~\ref{fig:ncdep} shows the summary plot for
$\chi/\sigma_{\rm str}^2$ and $b_2$, including our results.
In this plot, we use
$T_c/\sqrt{\sigma_{\rm str}}=0.7091(36)$~\cite{Lucini:2003zr} to change
the normalization to $\chi/\sigma_{\rm str}^2$.
The solid lines shown in the plots are the linear fit performed in
Ref.~\cite{Bonati:2016tvi} using the data at $N=3$, 4, 6.

The results of $\chi/\sigma_{\rm str}$ for SU(2) theory are slightly above
than the solid line, but the deviation is accountable by the next
leading order correction, which is of $O(1/N^2)$ relative to the leading
one.
It is then natural to expect that the dynamics of SU(2) gauge theory is a
smooth extrapolation of the large $N$ dynamics to $N=2$, and that nothing
special happens in between.

The value of $b_2$ at $N=2$ obtained in this work turns out to be consistent with the
instanton prediction, $b_2^{\rm DIGA}=-1/12$, within 1.7 $\sigma$.
However, it is more consistent with the naive linear extrapolation from
the $N\ge 3$ data to $N=2$.
This observation gives further support to the above expectation, {\it
i.e.} nothing special happens between $N\ge 3$ and $N=2$.
Notice that, in Ref.~\cite{Bonanno:2018xtd} $b_4=6(2)\times 10^{-4}$ is
obtained in the continuum limit, which clearly differs from the value predicted from the 
instanton calculus, $b_4^{\rm DIGA}=1/360$.

\section{Discussion}\label{sec:discussion}

\subsection{\texorpdfstring{Large $N$ versus Small $N$}{Large N versus Small N}}

One of the motivations for present analysis of the 4d $\SU(2)$ pure Yang-Mills theory is to 
study the $\SU(N)$ Yang-Mills theories for finite values of $N$.
We expect that $\SU(N)$ Yang-Mills theories show qualitatively different behaviors,
for large $N$ and for small $N$.

\subsubsection{\texorpdfstring{Large $N$}{Large N}}

\begin{figure}
\includegraphics[scale=0.7]{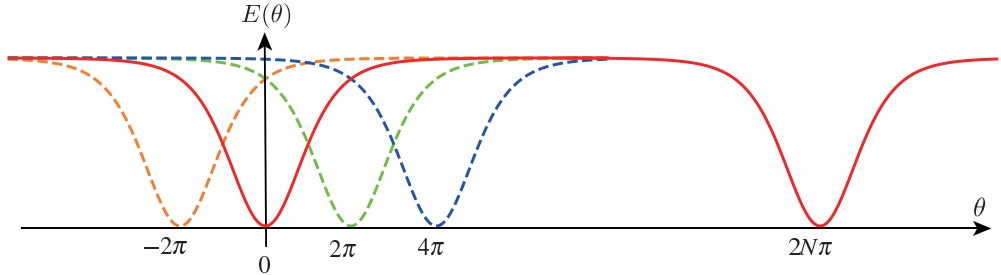}
\caption{Schematic picture for the multi-valued vacuum energy for the
 large $N$ pure Yang-Mills theory, reproduced from
 Ref.~\cite{Nomura:2017zqj}.}
\label{fig.branch}
\end{figure}

In the large $N$ limit, the values of the 
coefficients $b_{2i}$ scales at
$O(N^{-2i})$ \eqref{b_N}, and hence becomes smaller as $N$ becomes large.
This indicates that the vacuum energy,
$E(\theta, N)$, will no longer be a $2\pi$-periodic function of $\theta$.
While this seems to be in tension with the $2\pi$-periodicity of $\theta$ in the Lagrangian,
the apparent inconsistency is resolved by the possibility that
the vacuum energy $E(\theta, N)$ is a multi-valued function of $\theta$ \cite{Witten:1980sp,Witten:1998uka}.
Namely, there are multiple branches $\tilde{E}(\theta+2\pi n)$ labeled by an integer $n$,
where each $\tilde{E}(\theta)$ 
has a quadratic potential near $\theta\sim 0$ and has a plateau of height $O(\Lambda^4)$
as $\theta\to \pm \infty$, with $\Lambda$ being the dynamical scale.
The correct minimal energy (over all the branches) is then
\begin{align}\label{eq:E-theta-largeN}
E(\theta)= \min_{n\in \mathbb{Z}} \tilde{E}(\theta+2 \pi n)\;,
\end{align}
see Fig. \ref{fig.branch}.
This in particular means that there are two different lowest-energy
states for $\theta=\pi$ mapped each other under the CP symmetry, and
hence we expect CP to be spontaneously broken.

One should notice that the vacuum energy takes a different functional form 
as expected from the semiclassical one-instanton calculation
\cite{tHooft:1976snw}, which gives the free energy density as 
\begin{align}\label{Z_instanton}
E(\theta) \sim \int_0^{\infty}\! \frac{d\rho}{\rho^5} \, (\mu \rho)^{b_1} \exp\left[-\frac{8\pi^2}{g^2(\mu)} \right] (1-\cos \theta) \;.
\end{align}
Here $\rho$ is the size modulus of the instanton,
$g(\mu)$ is the running gauge coupling constant at the energy scale $\mu$,
and $b_1:=11N/3$ is the coefficient of the one-loop beta function.

One of the reasons for the discrepancy between the semi-classical instanton analysis 
and the large $N$ analysis resides in the famous IR divergence of the instanton analysis:
the integral \eqref{Z_instanton} is divergent as $\rho$ becomes large.

\subsubsection{\texorpdfstring{Small $N$}{Small N}}
\label{sec.small_N}

The situation can be different when $N$ is small.
Let us regard $N$ as a real parameter.
Then the integral \eqref{Z_instanton} has 
a UV divergence at $\rho\to 0$, if 
the values of $N$ is smaller than the threshold value 
$N_{\rm inst}=12/11$~\cite{Luscher:1981tq}.
In this case, the integral \eqref{Z_instanton} should be regularized in the UV by the cutoff scale $M$,
so that the lower value of the integral for $\rho$ is given by $M^{-1}$.
Note that the UV regularization is needed even though the 
pure Yang-Mills theory in itself is asymptotic free.

We thus expect that the one-instanton contribution to have the schematic
expression
\begin{align}\label{Z_M_Lambda}
E(\theta, N) \sim M^{4-\frac{11N}{3}} \Lambda^{\frac{11N}{3}} (1-\cos \theta) \;,
\end{align}
where $\Lambda$ is the dynamical scale of the theory.
This is a different qualitative behavior as suggested by the large $N$
analysis, \eqref{eq:E-theta-largeN}.
\footnote{While we discuss only pure Yang-Mills theory in this 
paper, similar issue arises for the $\SU(2)$ electroweak gauge group for
the standard model, and we need to introduce the UV regulator for the
small-size instantons. Interestingly, the size of the resulting integral
could explain the smallness of the cosmological constant
\cite{Nomura:2000yk,McLerran:2012mm,Ibe:2018ffn}.}

There can be several questions to this narrative.
First, in the analysis above for small $N$,
we evaluated only the instanton corrections, and one might object that 
there can be many other contributions to the partition function.
While this is certainly true, let us note that these non-instanton contributions give rise to contributions of 
$O(\Lambda^4)$.
Since the cutoff scale $M$ is much larger than the dynamical scale $\Lambda$ ($M\gg \Lambda$),
these non-instanton contributions are much smaller than the instanton contribution of 
\eqref{Z_M_Lambda}. Since we do know that instanton contributes to the 
path integral, we are certain that there is contribution of the form \eqref{Z_M_Lambda},
which we expect will dominate over other contributions.\footnote{There can be still multi-instanton corrections.
These subleading corrections preserve the $2\pi$-periodicity and hence the CP symmetry, and does not play important roles in
what follows.}

Note that the UV cutoff $M$ dependence will appear only as an overall divergence.
This means that while the topological susceptibility depends on the UV cutoff,
the coefficients $b_{2i}$ do not depend on the UV cutoff.
This is in fact an advantage of the definition of $b_{2i}$ in \eqref{E_expand}.

Another possible objection is that it does not make sense to 
consider non-integer values of $N$; $\SU(N)$ theory in the conventional thinking is defined only for integers $N\ge 2$.
For our purposes, however, it is useful to promote $N$ to be a real parameter
and discuss the vacuum energy $E(\theta,N)$ as a function of real values of $N$ as well as $\theta$.
Mathematically, one might worry that there are huge ambiguities in 
extending the functions $E(\theta,N)$ to non-integer values. 
Indeed, when we multiply the vacuum energy by an expression $F(\sin \pi N)$ for any function $F(x)$
with $F(x=0)=1$, the integer values of $E(\theta, N)$ will be preserved.
This however changes the asymptotic behavior as $N\to \infty$.
There is a mathematical theorem \cite{Carlson} which guarantees that 
two real-valued functions, with suitable asymptotic conditions at infinity 
and with the same values at all integers, coincide.
Such considerations are actually implicit in the large $N$ analysis,
and makes it possible to discuss small non-integer values of $N$ (even to $N<2$).
The threshold value $N_{\rm inst}$ for the instanton calculus makes sense in this context.

\subsubsection{\texorpdfstring{Intermediate $N$}{Intermediate N}}

We have seen that $\chi\sim O(N^0)$ and $b_{2i}\sim O(N^{-2i})$ for
large $N$ while $\chi$ depends on UV-cutoff and $b_{2i}\sim b_{2i}^{\rm
DIGA}$ for small $N$.

What happens at intermediate values of $N$?  For generic values of
$\theta$ we do not necessarily expect a sharp transition between ``large'' and
``small'' $N$: there is no good order parameter.  The
situation is different for the special value of $\theta=\pi$, which has
the CP symmetry in the Lagrangian.  In this case, we can define two
phases by the presence or the absence of the CP symmetry; the CP
symmetry is spontaneously broken for large $N$, while for smaller $N$
the vacuum energy may be given by the cosine form and hence CP is
preserved. In that case, there exists a critical value of
$N=N_{\rm CP}$ between the two phases.

\begin{figure}
\includegraphics[width=16cm]{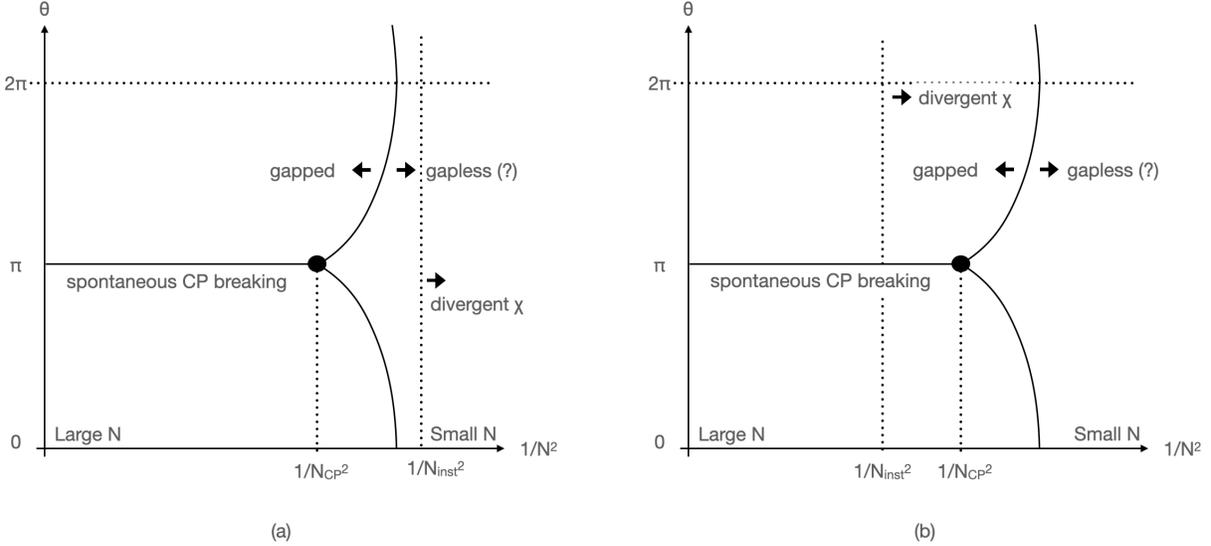}
\caption{ Possible phase structures of 4d SU($N$) pure Yang-Mills theory
as a function of $1/N^2$ and $\theta$. At $\theta = \pi$ and $N > N_{\rm CP}$, 
there is a first-order phase transition. If such $N_{\rm CP}$
exists, a gapless theory should be realized for $N < N_{\rm CP}$ at
$\theta = \pi$. Such a region may extends to $\theta \neq \pi$ as in the
figures although it is totally unknown.
The topological susceptibility, $\chi$, diverges for $N < N_{\rm inst} =
12/11$ by the contributions from small instantons.  In (a) and (b)
we show possible phase structures where $N_{\rm inst} < N_{\rm CP}$ and
$N_{\rm inst} > N_{\rm CP}$, respectively.  The mixed anomaly in itself
allows for more complicated phase structures, and for example allows regions where both CP and center symmetries are broken.}  
\label{fig.phase_pi}
\end{figure}

We expect that the two phases are different also in that whether the
vacuum is gapped or gapless.  This is because of the mixed anomaly
between the $\mathbb{Z}_N$ center symmetry and the CP symmetry
\cite{Gaiotto:2017yup}, and the presence of the $\mathbb{Z}_N$ symmetry
can be regarded as a definition of the confinement. Possible phase
diagrams are shown in Fig.~\ref{fig.phase_pi}, where the
presence/absence of the CP symmetry is assumed to coincide with the
gapped/gapless system.
Of course, the presence of the mixed anomaly only shows that at least
either the center symmetry or the CP symmetry should be broken, and
allows for the possibility that both are broken.

Here, the existence of $N_{\rm CP}$ is our assumption motivated by the
phase structure of the \cpn model we discuss below. Once it is
assumed, we need to discuss how the gapless phase extends to the
$\theta \neq \pi$ region. In the figure, we show a possibility that the
gapless theory is realized even at $\theta = 0$ at some $N$. There are
other possibilities that the line does not reach to $\theta = 0$ axis,
as well as the possibility that gapless theories are realized only on the
$\theta = \pi$ line.
Note that irrespective of the possible phase structures we define the
critical value $N_{\rm CP}$ by the presence/absence of the CP symmetry.

Note that the value of $N_{\rm inst}$ is not necessarily the same as 
the critical value $N_{\rm CP}$; the former is defined purely for the 
semiclassical instanton computation applicable for generic values of $\theta$,
while the critical value $N_{\rm CP}$ is the value 
separating the CP broken/preserved phases at the special value $\theta=\pi$.
It is not a priori clear if we expect general inequalities
between the two values $N_{\rm inst}$ and $N_{\rm CP}$.
One may be tempted to imagine that $N_{\rm inst}$ should always be smaller than 
$N_{\rm CP}$ since the potential generated by the instanton is always smooth
so that the spontaneous CP breaking does not happen. However, although
it is certain that the contributions from the small instantons dominate
the instanton density for $N < N_{\rm inst}$, one cannot exclude the
possibility that non-trivial infrared physics still leads the
spontaneous CP breaking and/or confinement at $\theta = \pi$.

Let us next come to more quantitative aspects.
In the instanton calculus, we obtained the threshold value of
$N_{\rm inst}=12/11$, which is smaller than $N=2$. 
The estimation by the one-loop beta function is justified by the
asymptotic freedom.
Therefore, it is expected that the $\SU(2)$ gauge theory has a
UV-independent value of the topological susceptibility. As we discussed
already, however, one cannot conclude whether $N_{\rm CP} < 2$ holds or not only
from this discussion. Below, we examine the lattice results of the
$\theta$-dependence of the theory and discuss whether $N=2$ is small or
large more carefully.

\subsubsection{
   \texorpdfstring{Comparison with the \cpn Model}
   {Comparison with the CP(N) Model}}

It is useful to compare the 4d Yang-Mills theory
with the celebrated \cpn model in two dimensions~\cite{Eichenherr:1978qa,DAdda:1978vbw}.
This theory has many similarities with the 4d SU($N$) Yang-Mills theory,
and could be of help in understanding the non-perturbative properties of
the latter.\footnote{In 2d \cpn a confining linear potential appears
even when instanton dominates the
dynamics~\cite{Luscher:1978rn,KeithHynes:2008rw}.}

In the large $N$ limit of the 2d \cpn model, there exists gap at any values of $\theta$,
and the vacuum energy is discontinuous at $\theta=\pi$, where the  CP
symmetry is spontaneously broken.
There seems to be a consensus that these properties takes over down
to $N=3$.
The situation is different from the 
$N=2$ case, {\it i.e.} the CP$^1$ model, which is nothing but the $\mathrm{O}(3)$ spin
model. This model is believed to
be gapless and have continuous vacuum energy at
$\theta=\pi$~\cite{KeithHynes:2008rw}.

When we consider $N$ as a continuous parameter again, 
one expects that there will be a critical value of $N$ between the two phases, which we denote by $N_{\rm CP}$. For $N>N_{\rm CP}$ the theory has 
spontaneous CP breaking at $\theta=\pi$, and we expect that the dependence on the parameter $N$
is accounted by the large $N$ scaling: we call this the ``large $N$ phase.''
By contrast for $N<N_{\rm CP}$ we have an unbroken CP symmetry for $\theta=\pi$,
and the semiclassical instanton analysis applies: we call this the ``small $N$ phase.''
The computation similar to \eqref{Z_instanton}
gives the threshold value $N_{\rm inst}=2$ for the \cpn model,
consistent with the divergence of topological susceptibility for the CP$^1$-model.

The most remarkable difference of the CP$^1$ model from other ($N>2$) \cpn models is that
the semi-classical calculation of the former leads to a UV divergence in the topological
susceptibility. This is supported by lattice numerical calculations,
unless a suitable counter term is added~\cite{Berg:1981er,Berg:1981nw,
Luscher:1981tq,Farchioni:1994fd,Blatter:1995ik,DElia:1995wxi,
Burkhalter:2001hu,Bietenholz:2010xg,Bietenholz:2018agd}\footnote{See
also Ref.~\cite{Berni:2020ebn}, in which the divergence of the
topological susceptibility is examined in detail.}.

\subsection{Quantitative Analysis of Lattice Results}
\label{subsec:4dsu2}

\subsubsection{\texorpdfstring{$N=2$}{N=2}}

We have already seen that the 
semi-classical estimate of the topological susceptibility  $\chi$ in 4d $\SU(N)$ gauge theory
does not yield UV divergence even for the possible smallest value,
$N=2$.
The continuum limit of lattice numerical calculations serves as an
independent quantitative test of this expectation.

Although an extrapolation of numerical data is always subtle, and
especially the continuum limit of quantities related to topological
charge needs special care, 
our result as well as previous results in the literature demonstrate the
finiteness of $\chi$ for SU($N$) gauge theory, all the way to the value
$N=2$.

As shown in Fig.~\ref{fig:ncdep}, the magnitude of $|b_2|$ obtained for
$N=2$ is slightly smaller than that of the instanton prediction
$b_2^{\rm DIGA}=-1/12$. The value of $|b_4|$ in
Ref.~\cite{Bonanno:2018xtd} is much smaller than the instanton value.
Both are rather consistent with the $1/N^2$ and $1/N^4$ scalings of the
$N \geq 3$ data as we discuss further below.  This suggests the
invalidity of the instanton description.
One also expects that the vacuum energy has a cusp at $\theta=\pi$ due
to small values of $|b_2|$ and $|b_4|$.
All these results suggest that $N=2$ is ``large'' for the
four-dimensional Yang-Mills theory.

In the literature there have been some attempts to analyze the vacuum of
the 4d $\SU(2)$ theory at $\theta=\pi$. For example, Ref.\
\cite{Unsal:2012zj} analyzes the question for a suitable double-trace
deformation of the 4d Yang-Mills theory via the semiclassical analysis
and a twisted compactification, and obtained the results consistent with
ours.  It should be kept in mind, however, that any deformation of the
theory, often needed for the semiclassical analysis, could potentially
change the vacuum structure of the theory, let alone the precise values
of $N_{\rm inst}$ and $N_{\rm CP}$. It is also the case that for our
discussion it is crucial to discuss the transition between
small $N$ and large $N$ behaviors, as we will discuss a few paragraphs
below.

Our result should be contrasted with the case of the 2d \cpn model,
where $N=2$ case is gapless and CP preserving at $\theta=\pi$, as already mentioned before.
This is an excellent demonstration of the quantitative differences between 
four-dimensional Yang-Mills theory and the two-dimensional \cpn model.

Notice that the relation between the 4d $\SU(N)$ Yang-Mills theory and
the 2d \cpn model was further clarified in \cite{Yamazaki:2017ulc},
which showed that the $T^2\times S^1$ compactification of the former
with suitable 't Hooft magnetic flux gives rise to $S^1$
compactification of the two-dimensional sigma model whose target space
has the topology of \cpn (see Refs.~\cite{Yamazaki:2017dra,Wan:2018zql}
for further checks via anomalies).
A caution is needed, however, before any quantitative comparisons between the two. The two-dimensional model obtained from four-dimensional theory
has a non-standard metric, and in addition there are special points (fixed points under the Weyl group action) 
in the \cpn where we encounter W-bosons of the four-dimensional theory \cite{Yamazaki:2017ulc}.
Moreover for the analysis of \cite{Yamazaki:2017ulc} it was crucial to have a hierarchy of scales between the sizes of 
$T^2$ and $S^1$, and any discussion of the standard flat space limit (where there is no such hierarchy) requires 
careful analytic continuation. These subtleties can easily affect quantitative discussions here.

\subsubsection{\texorpdfstring{$N_{\rm inst}$}{N(inst)}}

Once we are settled with the case of $N=2$, we can discuss even smaller
values of $N$ and ask how the theory approaches the $N < N_{\rm inst}$ region.

We first pretend that $N_{\rm inst}$ is unknown and try to determine its value by
the lattice data by two methods.
The first method uses topological susceptibility, which we expect to diverge
at the value $N=N_{\rm inst}$. To determine this value 
we fit our $N=2$ result together with those for $N=3$, 4, 6
in Ref.~\cite{Bonati:2016tvi} by an Ansatz
\begin{align}
 \frac{\chi}{\sigma_{\rm str}^2}
=\left.\frac{\chi}{\sigma_{\rm str}^2}\right|_{N=\infty}\times
\frac{N^2}{N^2-N_{\rm inst}^2}\ ,
\end{align}
where $N_{\rm inst}$ is assumed to be a real number.
Here the Ansatz is the simplest function of $N^2$ which has 
divergence at $N=N_{\rm inst}$ and approaches to the 
large $N$ value as $N\to \infty$.\footnote{Our Ansatz is motivated 
by the analysis performed in Ref.~\cite{Bonanno:2018xtd},
where the topological susceptibilities of 2d \cpn model are calculated
at several values of $N$ and fitted to the function including $1/(N-2)$.
Our overall conclusion is qualitatively unchanged if we modify this Ansatz. 
Note that the possible data points are rather limited, since 
we have only $N=2,3, \dots$, and all the points for $N\ge 3$ are already well-fitted by the large $N$ scaling.}
We then obtain
\begin{align}
\left.\frac{\chi}{\sigma_{\rm str}^2}\right|_{N=\infty}
= 0.0214(2)\ ,
\qquad
N_{\rm inst} = 1.52(10)\ \qquad \mbox{with $\chi^2/{\rm dof}=0.05$}
\label{eq:ncrit}
\end{align}
which is shown as the dashed curve in Fig.~\ref{fig:ncdep} denoted by
``ph fit'' meaning phenomenological fit.

The second method uses the values of $b_2$.
Supposing that the semi-classical calculation becomes valid at 
$N=N_{\rm inst}$ for SU($N$) gauge theory, $b_2$ is expected to take
$b_2^{\rm DIGA}=-1/12$ at the same value of $N$.
We again use the results for $b_2$ for $N=2$, 3, 4, 6 to test this
expectation.
This time, by fitting the data to
\begin{align}
 b_2(N)=\frac{b_2^{(1)}}{N^2}\ ,
\label{eq:b2-fitform}
\end{align}
we obtain 
\begin{align}
b_2^{(1)}= -0.200(12)\ \qquad \mbox{with $\chi^2/{\rm dof}=0.97$}.
\label{eq:b2-fit}
\end{align}
Substituting (\ref{eq:ncrit}) and (\ref{eq:b2-fit}) into 
(\ref{eq:b2-fitform}) yields $b_2(N_{\rm inst})=-0.087(5)$, which is
consistent with $-1/12$. 
Furthermore, assuming the functional form of $b_4(N)=b_4^{(1)}/N^4$ and
using the result $b_4=6(2)\times 10^{-4}$ at
$N=2$~\cite{Bonanno:2018xtd}, $b_4(N_{\rm inst})=0.0018(6)$ is 
obtained, which is not that far from $b_4^{\rm DIGA}=1/360$. 

The two methods produce consistent estimates $N_{\rm inst} \sim 1.5$,
slightly larger than the semi-classical value $12/11$. 
This numerology serves as a check of the overall picture, and moreover
indicates that the large $N$ scaling of $b_2$ holds
well all the way until the value $N\sim N_{\rm inst}$,
where $\chi$ diverges.
If we assume large $N$ scaling for all the $b_{2n}$'s until $N\sim N_{\rm inst}$,
then certain derivatives of the free energy are necessarily discontinuous
at $\theta=\pi$, thus implying the breaking of the CP symmetry. 
This suggests the inequality $N_{\rm CP} \lesssim N_{\rm inst}$.
Further numerical studied are needed to make this inequality more precise.

Summarizing our discussion,
the numerical data suggests the following shape of the vacuum energy density, as we change the value of $N$.
At large $N$ we have the quadratic form of the vacuum energy around $\theta=0$, while there is a cusp at $\theta = \pi$. 
As we change $N$ to smaller values, $b_2$ and $b_4$ grow while continue to obey the large $N$ scaling to a good approximation.
The cusp of the vacuum energy at $\theta=\pi$ is gradually smoothened, however not completely;
CP is still spontaneously broken. The transition of the large $N$ picture to the instanton picture seems to be smooth as far as $b_{2n}$ are concerned,
and $N=2$ is on the ``large $N$'' side.
At $N_{\rm inst} \sim 1.5$ the free energy approaches the cosine function with a diverging
overall factor, $\chi$. 
Once $\chi$ is diverging, it becomes difficult to
infer the vacuum structure from the vacuum energy only,
since the vacuum energy is masked by the large 
contributions of small instantons. In particular, it becomes invisible in practice whether or not there is a
phase transition at $\theta = \pi$.

\section{Summary and Discussion}\label{sec:summary}

We performed lattice numerical simulations to explore the $\theta$
dependence of the vacuum energy in 4d $\SU(2)$ pure Yang-Mills theory, with special
attention to the response of topological excitation to the smearing
procedure.
We discussed the method to extract the topological information from smeared
configurations properly and estimated the first two coefficients in the
$\theta$ expansion of the vacuum energy in the continuum limit, namely $\chi$ and $b_2$.
The value of $\chi$ turns out to be consistent with the previous results in the literature,
while $b_2$ is determined for the first time.

We use these results to infer the phase structure of the 4d $\SU(N)$ theory
as we change the values of $N$ and $\theta$.
We highlighted the differences for ``large $N$'' and for ``small $N$'':
we have the large $N$ scaling for the former, while the vacuum energy is dominated by instantons in the latter.
The differences between the two is most clear-cut for $\theta=\pi$,
where the CP symmetry is spontaneously broken for large $N$,
while unbroken for small $N$.

We found that for $N=2$ the topological susceptibility $\chi$ remains finite, and $b_2$
slightly deviates from the instanton predictions, while it is well fitted by the
$1/N^2$ extrapolation from the $N \geq 3$ values.
By further extrapolating $N$ to small values by analytic continuation, we find that $\chi$ and
$b_2$ reach the instanton predictions at $N_{\rm inst}=1.52(2)$,
and that the free energy
will be dominated by instantons.

Our analysis gives strong quantitative evidence that the $\SU(2)$ theory, and hence all $\SU(N)$ theories for integer $N$, are in the large $N$ category. 
While large $N$ analysis is often regarded as an approximation applicable only to the large values of $N$,
our results suggest that the large $N$ analysis is more powerful,
and can be useful for studying all possible values of $N$, even as small as $N=2$.
This is in contrast with the case of the 2d \cpn model,
which is believed to be gapless for $N=2$, while are gapped for $N\ge 3$.
It would be interesting to study for more general theories
the applicability of large $N$ analysis to smaller values of $N$.

In this work, we could not explore the question of precisely what topological object carries
non-zero topological charges.
In Ref.~\cite{Horvath:2003yj}, it was pointed out that in SU(3)
Yang-Mills theory the codimension-one objects (``sheets'') are responsible for the non-zero topological charges,
and from the similar study of \cpn model it was pointed out that the object becomes
localized and becomes instantons as $N\to N_{\rm CP}$.
Thus, it is interesting to see what objects are responsible for
the topological charges in the 4d $\SU(2)$ theory.

It is also interesting to see the $\theta$ dependence of the vacuum
energy directly on the lattice, for finite real values of $\theta$, especially near $\theta=\pi$.
This program has to overcome notoriously difficult problem, the sign
problem.
Since recent development in methodology is
remarkable~\cite{Hirasawa:2020bnl,Gattringer:2020mbf,Sulejmanpasic:2020lyq},
such direct studies appear to be within reach in the near future.

Finally, it is interesting to ask if the analysis of this paper 
has any phenomenological considerations of the dynamical $\theta$-angle,
the axion
\cite{Peccei:1977hh,Peccei:1977ur,Weinberg:1977ma,Wilczek:1977pj}.
For example, in the axionic inflationary models of
Ref.~\cite{Nomura:2017ehb} (see also Ref.~\cite{Dubovsky:2011tu}), the
values of $b_{2n}$ affect future observations of primordial
gravitational waves from inflation \cite{Nomura:2017zqj}.

\section*{Acknowledgments}

We would like to thank Hideo Matsufuru and Julien Frison for the support
on developing the codes used in this work, and Aleksey Cherman for discussion.
This work is in part based on Bridge++ code
(see, for details, http://bridge.kek.jp/Lattice-code/ and
Ref.~\cite{Ueda:2014rya}), and supported by JSPS KAKENHI Grant-in-Aid
for Scientific Research (Nos.~19H00689 [RK, NY, MY], 18K03662 [NY],
19K03820 [MY]) and MEXT KAKENHI Grant-in-Aid for Scientific Research on
Innovative Areas (No.~18H05542 [RK]).
Numerical computation in this work was carried out in part on the
Oakforest-PACS and Cygnus under Multidisciplinary Cooperative Research
Program (No.~17a15) in Center for Computational Sciences, University of
Tsukuba; the Yukawa Institute Computer Facility, Yukawa Institute, Kyoto
University; a supercomputer (NEC SX-5) at Research Center for Nuclear
Physics, Osaka University; Fujitsu PRIMERGY CX600M1/CX1640M1
(Oakforest-PACS) in the Information Technology Center, The University of
Tokyo.

\end{document}